\documentclass[fleqn,10pt]{wlscirep}
\usepackage[utf8]{inputenc}
\usepackage[T1]{fontenc}
\usepackage{float}
\usepackage{url}
\usepackage{multirow}
\usepackage{graphicx}
\usepackage{amsmath,amssymb,amsfonts}
\usepackage{lineno}

\begin{document}
\title{From Pretraining to Privacy: Federated Ultrasound Foundation Model with Self-Supervised Learning}

\author[1,2,3,$\dagger$]{Yuncheng Jiang}
\author[4,$\dagger$]{Chun-Mei Feng}
\author[1,2]{Jinke Ren}
\author[5]{Jun Wei}
\author[1,2]{Zixun Zhang}
\author[7]{Yiwen Hu}
\author[8]{Yunbi Liu}
\author[1,2]{Rui Sun}
\author[9]{Xuemei Tang}
\author[10]{Juan Du}
\author[6]{Xiang Wan}
\author[11]{Yong Xu}
\author[12]{Bo Du}
\author[13,14]{Xin Gao}
\author[15]{Guangyu Wang}
\author[16,17]{Shaohua Zhou}
\author[2,1]{Shuguang Cui}
\author[2,1,$\ddagger$]{Zhen Li}
\affil[1]{FNii-Shenzhen, The Chinese University of Hong Kong, Shenzhen, Shenzhen 518172, China}
\affil[2]{School of Science and Engineering, The Chinese University of Hong Kong, Shenzhen, Shenzhen 518172, China}
\affil[3]{West China Hospital of Sichuan University, Chengdu 610041, China$\ast$}
\affil[4]{School of Computer Science, University College Dublin, Ireland}
\affil[5]{College of Computer Science and Software Engineering, Shenzhen University, Shenzhen 518060, China}
\affil[6]{Shenzhen Research Institute of Big Data, Shenzhen 518172, China}
\affil[7]{South China Hospital, Health Science Center, Shenzhen University, Shenzhen 518111, China}
\affil[8]{School of Computer Science, Nanjing University of Posts and Telecommunications, Nanjing 210023, China}
\affil[9]{Affiliated Hospital of North Sichuan Medical College, Sichuan 637000, China}
\affil[10]{North Sichuan Medical College, Sichuan 637000, China}
\affil[11]{Bio-Computing Research Center, Harbin Institute of Technology, Shenzhen, Shenzhen 518055, China}
\affil[12]{School of Computer Science, Wuhan University, Wuhan 430072, China}
\affil[13]{Computer Science Program, Computer, Electrical and Mathematical Sciences and Engineering Division, King Abdullah University of Science and Technology (KAUST), Thuwal 23955-6900, Kingdom of Saudi Arabia.}
\affil[14]{Center of Excellence for Smart Health (KCSH), King Abdullah University of Science and Technology (KAUST), Thuwal 23955-6900, Kingdom of Saudi Arabia.}
\affil[15]{Beijing University of Posts and Telecommunications, Beijing 100876, China}
\affil[16]{School of Biomedical Engineering, Suzhou Institute for Advanced Research, University of Science and Technology of China, Suzhou 215123, China 
}
\affil[17]{Institute of Computing Technology, Chinese Academy of Sciences, Beijing 100190, China}
\affil[$\dagger$]{The authors contributed equally to this work.}
\affil[$\ddagger$]{Corresponding authors: Zhen Li (e-mail: lizhen@cuhk.edu.cn)}
\affil[$\ast$]{The first author worked as a postdoctoral fellow at West China Hospital of Sichuan University after graduation}
% \keywords{Keyword1, Keyword2, Keyword3}

\begin{abstract}
Ultrasound imaging is widely used in clinical diagnosis due to its non-invasive nature and real-time capabilities. 
However, traditional ultrasound diagnostics relies heavily on physician expertise and is often hampered by suboptimal image quality, leading to potential diagnostic errors. While artificial intelligence (AI) offers a promising solution to enhance clinical diagnosis by detecting abnormalities across various imaging modalities, existing AI methods for ultrasound face two major challenges.  
\textit{First}, they typically require vast amounts of labeled medical data, raising serious concerns regarding patient privacy. \textit{Second}, most models are designed for specific tasks, which restricts their broader clinical utility. To overcome these challenges, we present \textbf{UltraFedFM}, an innovative privacy-preserving ultrasound foundation model. UltraFedFM is collaboratively pre-trained using federated learning across 16 distributed medical institutions in 9 countries, leveraging a dataset of over 1 million ultrasound images covering 19 organs and 10 ultrasound modalities. This extensive and diverse data, combined with a secure training framework, enables UltraFedFM to exhibit strong generalization and diagnostic capabilities. It achieves an average area under the receiver operating characteristic curve (AUROC) of 0.927 for disease diagnosis and a dice similarity coefficient (DSC) of 0.878 for lesion segmentation. Notably, UltraFedFM surpasses the diagnostic accuracy of mid-level ultrasonographers (4–8 years of experience) and matches the performance of expert-level sonographers (10+ years of experience)  in the joint diagnosis of 8 common systemic diseases.c These findings indicate that UltraFedFM can significantly enhance clinical diagnostics while safeguarding patient privacy, marking a significant advancement in AI-driven ultrasound imaging for future clinical applications.
\end{abstract}
% \linenumbers

\flushbottom
\maketitle
\thispagestyle{empty}

\section*{Introduction}
\label{introduction}
% 超声图像应用现状
% 传统方法的局限
Ultrasound is becoming increasingly important in clinical practice worldwide. It offers significant advantages over magnetic resonance imaging (MRI) and computed tomography (CT), including freedom from radiation, non-invasive nature, and cost-effectiveness. Thus, it is widely adopted as the primary imaging method for monitoring fetal growth during pregnancy \cite{whitworth2015ultrasound}, diagnosing internal organ pathology, and assisting in surgical decision-making \cite{akkus2019survey}. However, ultrasound-based diagnosis relies heavily on the clinician's experience, while factors like noise and artifacts in the images can compromise quality and hinder the clinician's assessment of pathological regions, increasing the risk of missed or incorrect diagnoses \cite{donofrio2014diagnosis, feldman2009us}. {Recent efforts have turned to artificial intelligence (AI) technologies to mitigate ultrasound-specific artifacts (e.g. speckle, false textures) and enhance diagnostic accuracy \cite{buv, fetal_plane_1, gbcu, yadav2022despeckling, yadav2023objective, yadav2024deep, yadav2024machine, yadav2024systematic, yan2025development, virmani2019assessment, dass2020image}. These contributions demonstrate that careful preprocessing, task-specific network design, and curated annotations can substantially improve performance for single-organ tasks.} Despite notable successes, existing AI-based ultrasound models typically focus on very specific medical scenarios and require large amounts of high-quality labeled data, which restricts their scalability and generalizability across diverse medical applications.

Over the past two years, foundational models (FMs) have attracted much attention due to their generality and high performance. In the medical field, many efforts \cite{USFM, EchoFM} have leveraged unlabeled ultrasound data to pre-train FMs and fine-tuned them for specific tasks using labeled data. However, existing ultrasound foundational models (USFMs) face three key challenges: (1) \textbf{Data privacy}. Ultrasound data are distributed across multiple medical institutions and cannot be shared due to privacy regulations (e.g., GDPR \cite{voigt2017eu}), restricting the volume of data available for pre-training; (2) \textbf{Limited modality}. Many USFMs are designed for particular ultrasound imaging modalities (e.g., echocardiograms), limiting their applicability to other imaging modalities and reducing their versatility; (3) \textbf{Imbalanced data distribution}. Existing USFMs often face an imbalance caused by the long-tailed distribution of the organ/lesion types represented in the dataset (e.g., 91\% breast ultrasound in 3M-US \cite{USFM}), leading to a biased performance in diagnosing uncommon conditions. These challenges highlight the need for new solutions that simultaneously address data privacy, scalability, and generalizability across various ultrasound imaging modalities and clinical scenarios.

\begin{figure}[H]
    \centering
    \includegraphics[width=\linewidth]{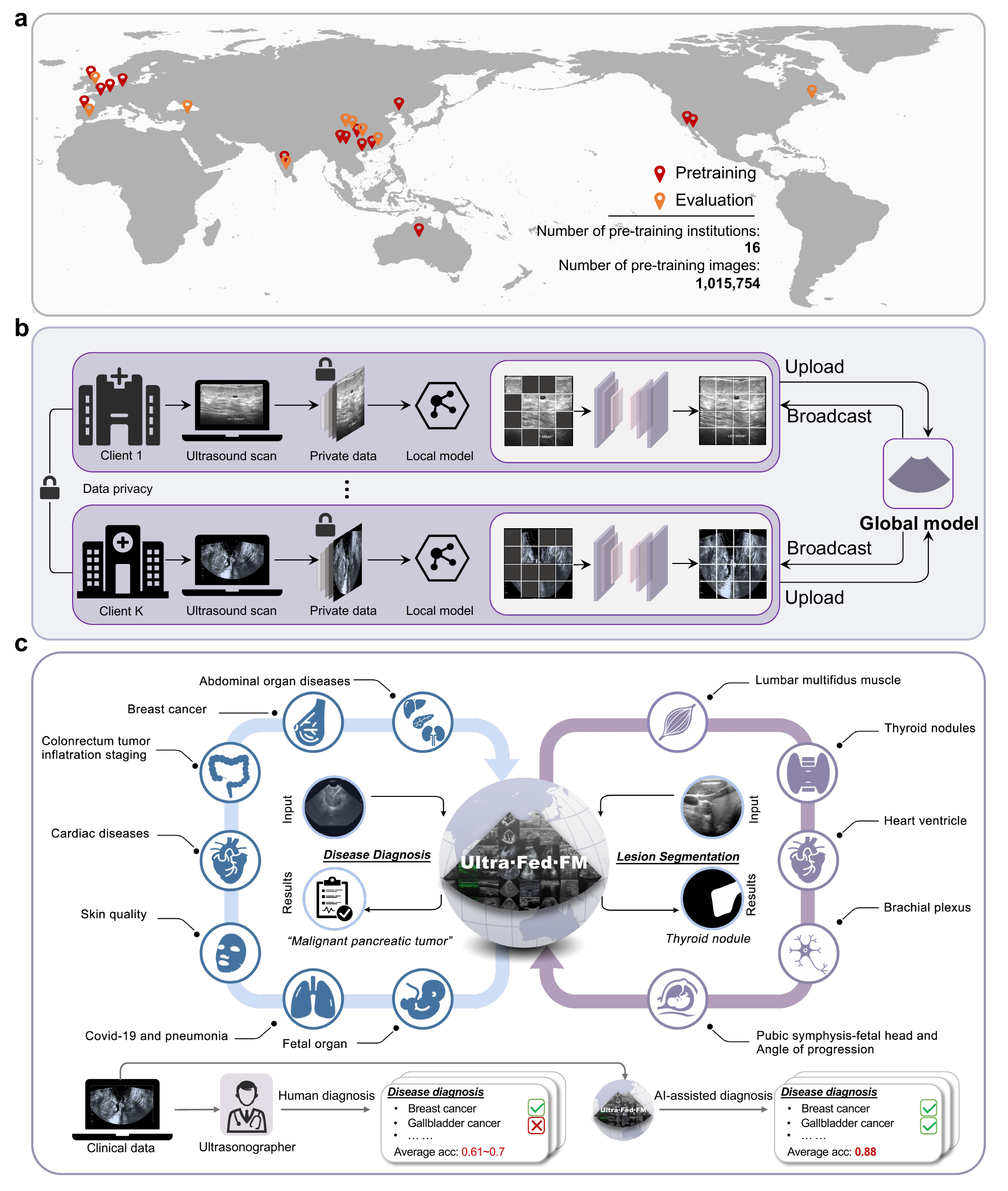}
    \caption{\textbf{Overview of the study.} \textbf{a} Medical data from 16 institutions and 9 countries are collected to pre-train and evaluate UltraFedFM, encompassing 1 million ultrasound images with extensive diversity. \textbf{b} The pre-training framework of UltraFedFM, where each client uses its private data to pre-train a local model through pixel-level reconstruction. During pre-training, only the local model parameters are uploaded for learning the global model, thus eliminating the risk of privacy breaches. Icons used are free to download from www.iconfont.cn and do not involve commercial use. \textbf{c} Clinical applications of UltraFedFM. UltraFedFM is a versatile ultrasound foundation model capable of handling multiple ultrasound scenarios, supporting multi-disease, multi-modal, and multi-task applications, and demonstrating superior performance compared with ultrasonographers in real clinical scenarios. Icons used are free to download from www.iconfont.cn and do not involve commercial use.}
    \label{fig:overview}
\end{figure}

\begin{figure}[H]
    \centering
    \includegraphics[width=\linewidth]{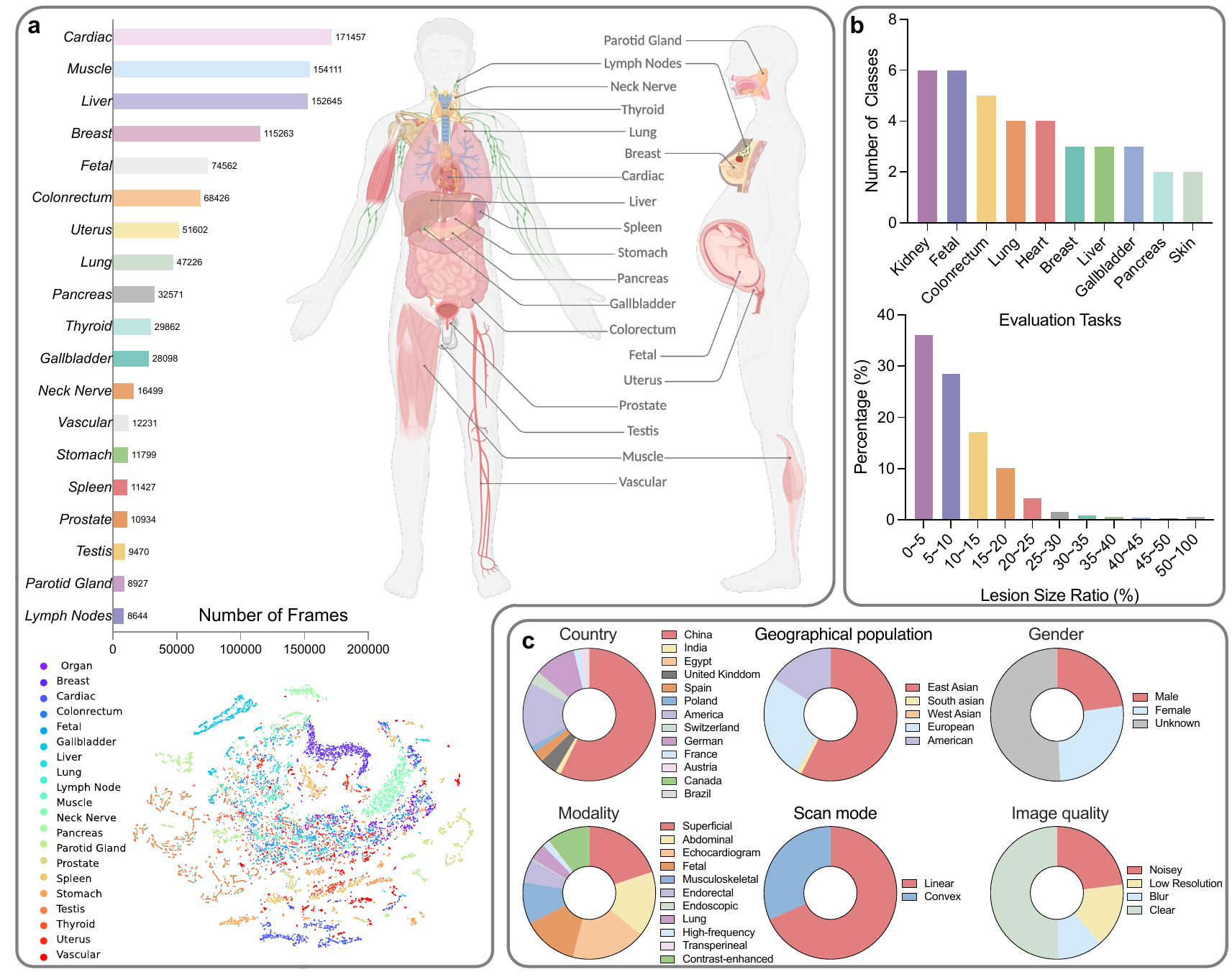}
    \caption{ \textbf{Statistics of the pre-training and downstream validation datasets.} \textbf{a} The pre-training dataset covers 19 major organs across the entire body captured by various ultrasound imaging modalities. The figure are created in Biorender and have obtained publication license. \textbf{b} The distribution of class numbers for each downstream diagnosis dataset, ranging from basic binary classification to complex multi-class classification, and the distribution of the target size for organ and lesion segmentation tasks in the downstream validation dataset. Most segmentation targets occupy less than 1/10 of the entire image. \textbf{c} The distribution of sensitive information in the dataset across six attributes.}
    \label{fig:dataset}
\end{figure}

% 我们的工作
In this work, we introduce UltraFedFM, a novel ultrasound foundation model pre-trained collaboratively by multiple medical institutions without exposing and aggregating all the data together. Specifically, we utilize a federated learning framework with one server and $16$ clients from $9$ countries, collectively possessing $1,015,754$ unlabeled ultrasound images (Fig. \ref{fig:overview}a). These images cover 19 systemic organs and 10 ultrasound imaging modalities ({Fig. \ref{fig:dataset}a}), providing an extensive and diverse representation for pre-training. By leveraging large-scale unlabeled datasets, UltraFedFM addresses key challenges in the medical field with the following solutions: \textbf{a.} When new modalities or organ data are continuously introduced, UltraFedFM can continuously update the model on new clients without accessing private data from other clients, thereby effectively safeguarding patient privacy; \textbf{b.} UltraFedFM minimizes the reliance on labor-intensive annotations by medical professionals, overcoming a critical bottleneck in medical AI development. This unsupervised approach ensures that valuable medical data can be efficiently utilized without requiring costly annotations from medical experts. The development of UltraFedFM consists of two stages: (1) Federated pre-training, in which the multiple clients collaboratively pre-train a shared model in a distributed, self-supervised manner. Throughout the pre-training process, the server periodically aggregates the local model parameters from each client without accessing their private data (Fig. \ref{fig:overview}b); (2) Downstream fine-tuning, where the pre-trained FM is fine-tuned using specific data to adapt to various clinical tasks, such as disease screening and diagnosis, sub-classification of disease phenotypes (e.g., tumor infiltration depth and type classification), prenatal maternal-fetal health analysis, and critical lesion identification and segmentation (Fig. \ref{fig:overview}c).

UltraFedFM is adapted to various ultrasound imaging modalities, modes, qualities, and clinical tasks. To accommodate the diverse features of different modalities, we propose a dynamic ultrasound image masking approach based on the specific texture features of organs and lesions. Additionally, we incorporate a random image corruption branch within the masked image modeling process to handle low-quality images commonly encountered in real-world scenarios. Furthermore, we use simple yet effective image transformations to generate simulated ultrasound images, aiming to address the uneven distribution of scan patterns in the pre-training dataset.

We conduct extensive experiments to evaluate the performance of UltraFedFM. To provide a fair and comprehensive evaluation, we collect and curate the largest ultrasound evaluation benchmark, covering the two most common ultrasound clinician tasks (i.e., disease diagnosis and lesion segmentation) with 11 sub-tasks from 19 ultrasound datasets. Several fully-supervised methods and a state-of-the-art USFM \cite{USFM} are utilized for comparison. Experimental results demonstrate that UltraFedFM outperforms all baselines, achieving an average area under the curve (AUROC) of 0.927 for disease diagnosis and a dice similarity coefficient (DSC) of 0.878 for lesion segmentation.
Notably, UltraFedFM outperforms ultrasonographer clinicians with intermediate levels (e.g., 4-8 years of clinical experience) and achieves comparable performance to high-level (e.g., more than 10 years of clinical experience) ultrasonographers in the joint diagnosis of 8 common systemic diseases. 
Furthermore, UltraFedFM leverages the principles of federated learning, enabling continuous model updates without the need for centralized data aggregation. This capability ensures the model to be further trained using private data from different institutions while preserving privacy and adhering to data protection regulations. By avoiding direct data sharing, UltraFedFM addresses critical privacy concerns and fostering collaboration across institutions. With these capabilities, UltraFedFM provides a reliable model and making it a pioneering solution for advancing ultrasound AI across institutions, regions, and clinical tasks.

\section*{Results}
\label{results}
% \subsection*{UltraFedFM enables systemic disease diagnosis}
{\subsection*{UltraFedFM enables systemic disease diagnosis and can assist clinicians in the diagnostic process}}
UltraFedFM aims to serve as a comprehensive FM in ultrasound imaging. To assess its effectiveness for disease diagnosis, 6 publicly available datasets and 2 private datasets (see Supplementary Table 2) are utilized, covering 8 kinds of organs (i.e., pancreas, gallbladder, liver, lung, colorectum, breast, heart, and fetal organs, see Supplementary Fig. 6) and 6 ultrasound imaging modalities (i.e., abdominal, lung, endorectal, superficial, echocardiogram, and fetal ultrasound). To provide an overall assessment for UltraFedFM, we average its performance on these datasets, and compare it with 4 baseline methods, including supervised training from scratch, ImageNet-21k centralized pre-training, USFM \cite{USFM} centralized pre-training, and masked autoencoder (MAE) \cite{MAE} federated pre-training. More details of the four methods are described in the Method section. The experimental results are shown in Fig. \ref{fig:diagnosis}. We observe that UltraFedFM achieves an average AUROC of 0.927, which significantly ($p<0.05$) outperforms its counterparts, surpassing the second-best USFM with an average AUROC of 0.894, by 0.033 ($p=0.002$). Additionally, UltraFedFM performs well in data-limited situations (see line plot in {Fig. \ref{fig:diagnosis}b}). As the amount of fine-tuning data is progressively reduced to 80\%, 60\%, 40\%, and 20\%, UltraFedFM remains robust, with only a modest decline in average AUROC of 0.124 (fine-tuning data from 100\% to 20\%), outperforming other methods. Notably, due to the federated pre-training with the large volume of unlabeled data, UltraFedFM possesses powerful feature extraction capabilities and can identify various types of lesions using a single organ-agnostic decoder, thus eliminating the need for task-specific classifiers utilized in other FMs. To demonstrate this, we constructed an organ-agnostic dataset by combining eight distinct datasets from different organs and fine-tuned UltraFedFM to recognize eight types of malignant tumors. It is observed that UltraFedFM accurately identifies most categories without requiring a separate classifier for each organ and the predicted scores of UltraFedFM concentrate in higher confidence intervals ({Fig. \ref{fig:diagnosis}c}). {Fig. \ref{fig:diagnosis}d} illustrates the receiver operating characteristic (ROC) curves among eight different diseases, showing that UltraFedFM achieves superior efficiency in organ-agnostic disease diagnosis. More quantitative results for UltraFedFM, including accuracy, F1-score, and ROC, are provided in Supplementary Fig. 7, Supplementary Fig. 8, and Supplementary Fig. 9.

{
To evaluate the reliability of UltraFedFM's generalist intelligence in clinical practice, we compare it with ultrasonographers having different clinical levels. Seven ultrasonographers participated in this study, of whom two are intermediate-level (clinicians A, B: 4-8 years of clinical experience) and five are high-level (clinicians C-G: more than 10 years of clinical experience). A total of 80 ultrasound images containing 8 systemic malignant diseases were tested. As shown in {Fig. \ref{fig:diagnosis}e} and Supplementary Table 7, UltraFedFM outperforms the ultrasonographers with intermediate-level and achieves comparable performance with high-level ultrasonographers. More specifically, while some specific organ diseases are easy for ultrasonographers to diagnose (e.g., average accuracy: 0.800 for breast and 0.871 for kidney), their diagnostic capabilities are limited when multiple ultrasound diseases are jointly diagnosed (e.g., average accuracy: 0.314 for gallbladder). In contrast, UltraFedFM can provide a consistent and accurate diagnosis of different ultrasound organ diseases (average accuracy: 0.900 for breast, 1.000 for kidney, and 0.800 for gallbladder. These results reveal that the UltraFedFM has the potential to serve as a reliable decision support tool to assist clinicians in prioritizing cases, reducing repetitive workloads, and minimizing missed diagnoses. 
}

\subsection*{UltraFedFM facilitates organ and lesion segmentation}
Organ and lesion segmentation for ultrasound images is crucial for clinical decision-making. To assess UltraFedFM's segmentation accuracy across different ultrasound imaging modalities, we evaluated it on four binary segmentation datasets (nerve \cite{nus}, muscle \cite{mus}, heart \cite{cardiacuda}, and thyroid \cite{thyroidcineus,ddti,tg3k,tn3k-1}) and one multi-class segmentation dataset (pubic symphysis-fetal head \cite{jnu}). UltraFedFM consistently achieves high segmentation accuracy, successfully managing targets with diverse shapes and structures. In the binary segmentation task (Fig. \ref{fig:segmentation}a), UltraFedFM achieves the highest average dice similarity coefficient (DSC) score of 0.857 across the three binary segmentation datasets, significantly outperforming all other baselines ($p<0.005$). In particular, USFM achieves a DSC score of 0.828, which is much lower than that of UltraFedFM ($p=0.002$).

The multi-class segmentation task involves two steps, beginning by segmenting the pubic symphysis and fetal head, followed by measuring the angle between them. In this task, UltraFedFM achieves a DSC score of 0.842, significantly outperforming the second-best method (USFM\cite{USFM}) with a DSC score of {0.810} ($p = 0.004$). Additionally, UltraFedFM excels in measuring the angle of progression (AoP), with a mean absolute error of 8.80, outperforming all baselines by a significant margin ($p < 0.005$) (Fig. \ref{fig:segmentation}a). Similar to the classification settings, we also evaluated UltraFedFM’s effectiveness in scenarios with limited labeled data ({Fig. \ref{fig:segmentation}b}). Notably, even with 20\% of the fine-tuning data, UltraFedFM still achieves an average DSC score of 0.772, outperforming the supervised method and USFM by 14.0\% and 2.3\%, respectively. To further assess UltraFedFM’s generalization capability, we compiled an organ-agnostic segmentation dataset comprising five types of lesions. As shown in {Fig. \ref{fig:segmentation}c}, UltraFedFM demonstrates superior performance in locating and segmenting these lesions using a single unified segmentation model.

We also conducted cross-institutional validation ({Fig. \ref{fig:segmentation}d}) and imbalanced scanning mode validation ({Fig. \ref{fig:segmentation}e}). The former explores the segmentation generalization capability of UltraFedFM across different organ modalities, while the latter evaluates its performance under varying scanning modes. Across all cross-validation datasets, UltraFedFM consistently outperformed other baseline models ($p<0.01$), demonstrating exceptional stability and balanced generalization capability. {Fig. \ref{fig:segmentation}e} shows the fine-tuning performance of UltraFedFM under different data ratios of linear array and convex array scanning modes. When the data distribution was highly imbalanced (e.g., 0\%:100\% or 100\%:0\%), both models exhibited uncontrollable bias and overfitting during training, leading to a decline in prediction performance. In contrast, when the data proportions were more balanced, the models achieved optimal performance. This indicates that the feature distribution of images plays a crucial role in both pre-training and fine-tuning stages.

{
\subsection*{UltraFedFM outperforms existing ultrasound task-specific methods}
To further evaluate the excellence of UltraFedFM in medical image analysis tasks, we comprehensively compared it with existing task-specific methods in the ultrasound field. We chose four representative tasks, namely fetal plane classification, gallbladder cancer classification, breast nodule segmentation, and thyroid nodule segmentation, for evaluation. For each task, we chose both traditional models and the latest high-performing methods as comparisons. Detailed information about the datasets and comparative methods is presented in the following Supplementary Table 8 and Supplementary Table 9. The results 

\begin{figure}[H]
    \centering
    \includegraphics[width=\linewidth]{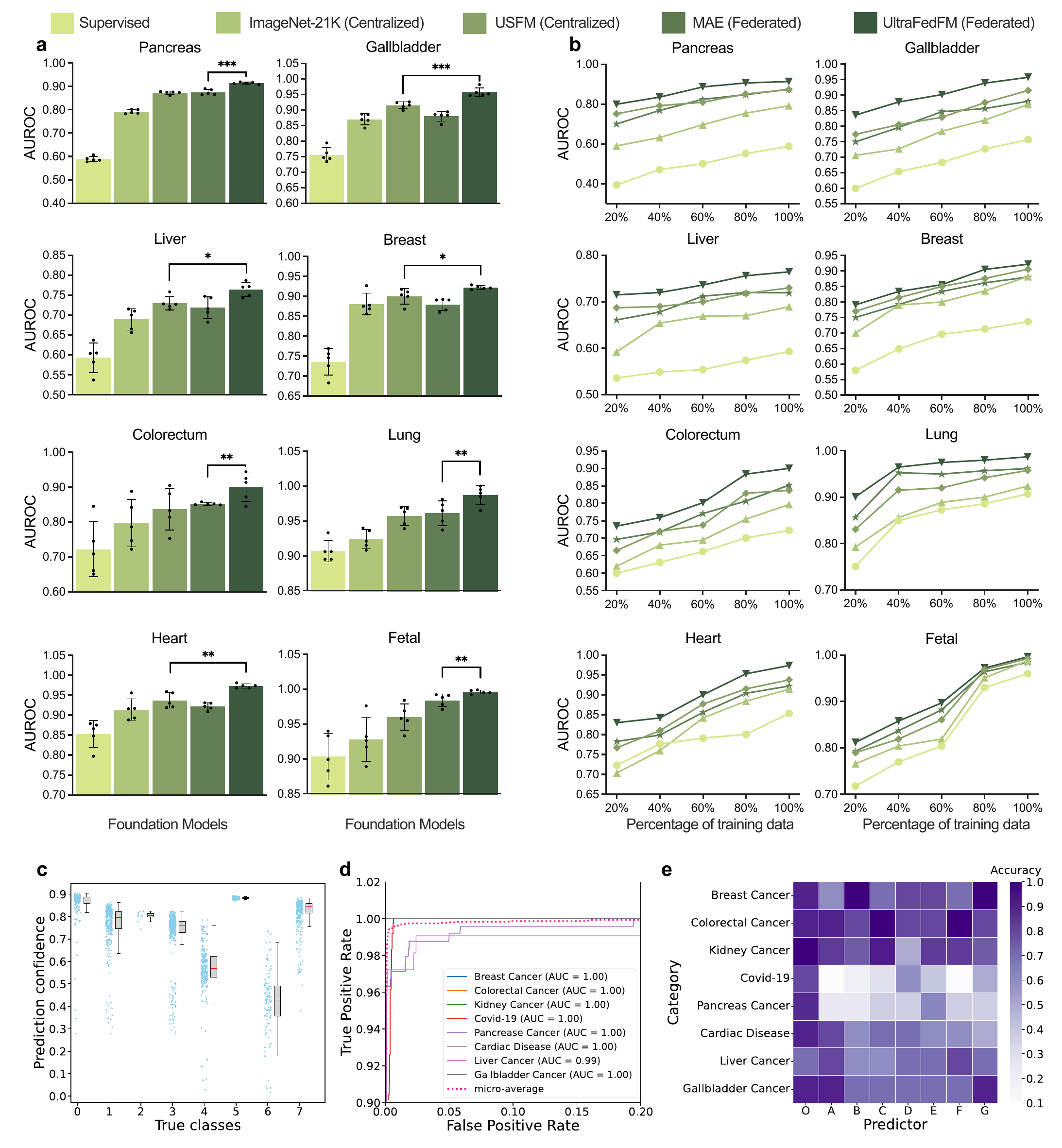}
    \caption{\textbf{Diagnostic performance for systemic disease classification.} \textbf{a} Internal validation of disease classification performance across eight diagnostic tasks. Comparative analysis shows model performance when fine-tuned on complete datasets. For each task, we fine-tune the model with five different random seeds. The error bars show $95\%$ confidence intervals (CI) of the estimates, and the bar center is the mean estimate. We compare the performance using the area under the receiver operating characteristic curve (AUROC). $P$-value is calculated with the two-sided t-test between UltraFedFM and the most competitive comparison model. $\ast$, $\ast\ast$, $\ast\ast\ast$ denotes $p<0.05$, $p<0.01$, and $p<0.001$. \textbf{b} The experimental results of disease classification on limited labeled data subsets. \textbf{c-d} Performance of UltraFedFM on organ-agnostic fine-tuning setting. \textbf{c} Prediction confidence distribution over eight disease classes. The center line of the box denotes the median, while the box edges represent the first and third quartiles and the whiskers extend to 1.5 times the inter-quartile range. \textbf{d} Receiver Operating Characteristic (ROC) curves for distinct disease categories. \textbf{e} Generalist diagnostic accuracy of UltraFedFM across eight diseases and the comparison with seven experienced ultrasonographers.}
    \label{fig:diagnosis}
\end{figure}

\begin{figure}[H]
    \centering
    \includegraphics[width=\linewidth]{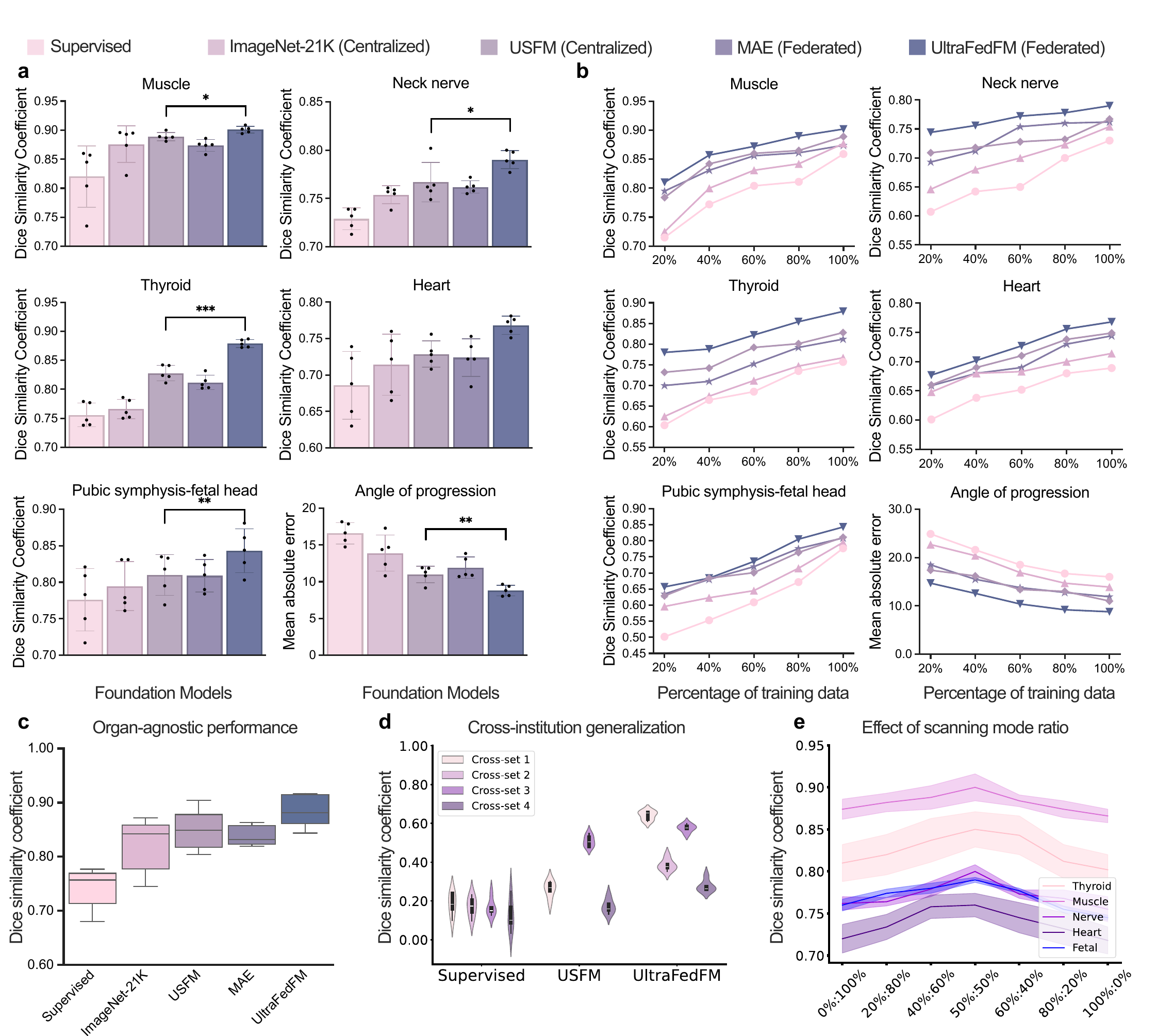}
    \caption{\textbf{Performance for organ and lesion segmentation.} \textbf{a} Internal validation of segmentation performance across eight diagnostic tasks. Comparative analysis shows model performance when fine-tuned on complete datasets. We compared the performance using the Dice similarity coefficient (DSC) score. \textbf{b} The experimental results of disease classification on limited labeled data subsets. \textbf{c} Performance of UltraFedFM on organ-agnostic dataset fine-tuned with a single decoder. \textbf{d} Comparison of cross-institution generalization performance. Cross-set 1 denotes fine-tuning on thyroid dataset and test on muscle dataset; Cross-set 2 denotes fine-tuning on thyroid dataset and test on muscle dataset; Cross-set 3 denotes fine-tuning on muscle dataset and test on thyroid dataset; Cross-set 4 denotes fine-tuning on muscle dataset and test on nerve dataset. The width represents the density of the data points at different values. The central line within each violin indicates the median.
    \textbf{e} Organ and lesion segmentation performance of UltraFedFM on different ratios of linear- and convex-array scanning mode ultrasound imaging data. The 95\% CI of DSC is plotted in color bands, and the center points of the bands indicate the mean value of DSC.
    }
    \label{fig:segmentation}
\end{figure} 

\noindent are illustrated in Fig. \ref{fig:generalization}a. The first two sub-figures display the results of classification tasks (evaluated by accuracy), while the last two sub-figures present the results of segmentation tasks (evaluated by the Dice similarity coefficient). Overall, UltraFedFM outperforms most of the comparative models in all evaluation metrics and tasks. In classification tasks, for the fetal plane classification task, UltraFedFM achieves an accuracy of 0.956, significantly higher than cutting-edge models such as HCN \cite{tanwar2024hcn} and Krishna et al. \cite{krishna2024standard} ($p<0.001$). On the gallbladder cancer classification dataset, it reaches an accuracy of 0.934, far surpassing classic models like ResNet \cite{he2016resnet} and Vision Transformer (ViT) \cite{vit}. Compared with the latest task-specific methods FocusMAE \cite{gbusv} and GBCHV \cite{hasan2025gbchv}, it shows improvements of 2.3\% and 1.9\% respectively ($p<0.001$). In segmentation tasks, on the breast nodule segmentation dataset, UltraFedFM obtains a Dice coefficient of 0.887, greatly outperforming advanced methods such as FABRFnet \cite{liu2025fabrfnet} and EMGANet \cite{huang2025emganet}, and achieving performance comparable to the state-of-the-art method nnU-Net \cite{isensee2021nnunet}. It is worth emphasizing that nnU-Net, as a standardized segmentation framework integrating data-adaptive processing and two-stage segmentation techniques, has achieved top-level performance in multiple segmentation tasks. On the thyroid nodule segmentation dataset, UltraFedFM achieves a Dice coefficient of 0.882, outperforming nnU-Net ($p<0.01$).
}

\subsection*{UltraFedFM generalizes to new medical scenarios}
Beyond learning ability, a crucial metric to evaluate the practicality of FMs in real-life scenarios is the generalization ability. To assess this, we selected two medical institutions not involved in the pre-training stage (high-frequency skin ultrasound imaging dataset and kidney disease ultrasound imaging dataset). This evaluation aims to determine how well the model performs on unseen ultrasound imaging modalities and organs, both key challenges in ultrasound diagnostics. As shown in {Fig. \ref{fig:generalization}b-c}. UltraFedFM consistently demonstrates superior generalization across different modalities, achieving an average AUROC of 0.925, significantly outperforming all other baselines ($p < 0.01$). {Fig. \ref{fig:visualization}c} illustrates that UltraFedFM achieves an AUROC of 97.1\% and an AP of 0.910, despite the textural and color differences of high-frequency ultrasound imaging from conventional methods. Such generalization is essential for real-world applications where clinicians frequently encounter new organs or modalities that the training data may not easily access.

\subsection*{The stability of UltraFedFM's predictions}
The stability of model predictions is essential for ensuring reliable clinical decision-making, particularly in ultrasound-based diagnostics, where inconsistencies can lead to misdiagnosis. To this end, we quantitatively compared the prediction stability of UltraFedFM with the baseline USFM \cite{USFM} under two settings: organ-specific (Fig. \ref{fig:discussion}a) and organ-agnostic (Fig. \ref{fig:discussion}b). USFM shows a broader distribution (mean $\mu = 0.808$ and standard deviation $\sigma = 0.174$). In contrast, UltraFedFM's predictions concentrate in a high DSC range (mean $\mu = 0.857$ and standard deviation $\sigma = 0.103$). 
Moreover, stability is paramount when dealing with organs that exhibit significant inter-patient variability, such as the liver or kidneys. Thus, we further introduce random noise to simulate real-world ultrasound imaging perturbations, such as tissue movement, operator variability, or imaging artifacts. We compared the test results under varying levels of noise. Despite these disturbances, UltraFedFM maintained highly correlated test scores ({Fig. \ref{fig:discussion}c}), demonstrating its robustness and reliability in clinical environments where imaging conditions can be unpredictable.

\subsection*{The scaling efficiency in UltraFedFM}
\begin{figure}[H]
    \centering
    \includegraphics[width=\linewidth]{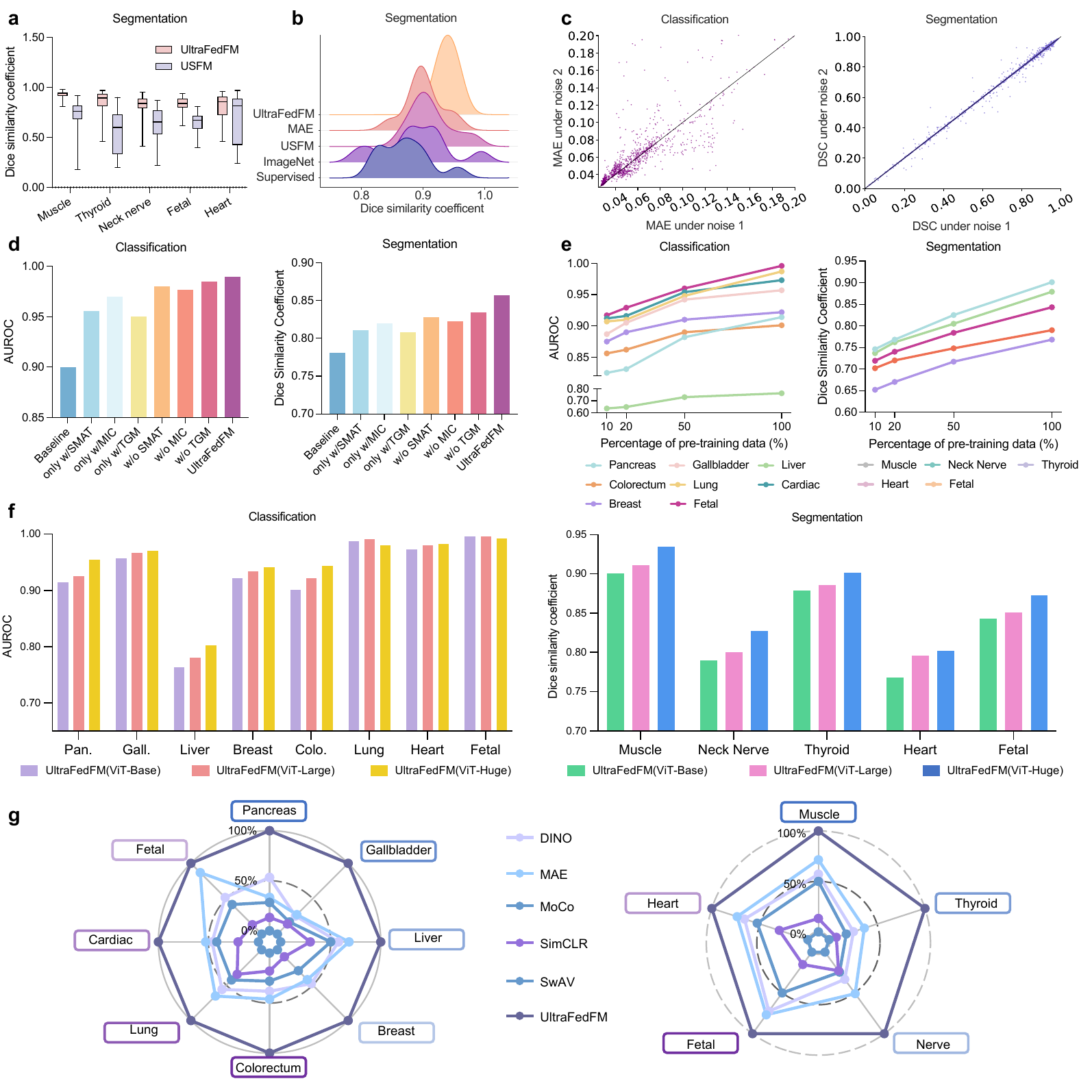}
    \caption{
    \textbf{a} Comparison of the prediction distribution between UltraFedFM and USFM across five independent segmentation tasks. UltraFedFM's predictions are concentrated within a high dice similarity coefficient range (mean $\mu=0.857$, standard deviation $\sigma = 0.103$), whereas USFM’s predictions show greater dispersion (mean $\mu=0.808$, standard deviation $\sigma = 0.174$) \textbf{b} Prediction distribution of all methods on organ-agnostic segmentation tasks. \textbf{c} The prediction stability of UltraFedFM under different ratios of input distribution variability. \textbf{d} The ablation study of proposed framework components. \textbf{e} The scaling effect of pre-training data, evaluated with different proportions of pre-training data. \textbf{f} The scaling effect of pre-training model size, evaluated using different ViT architectures (ViT-Base, ViT-Large, and ViT-Huge). \textbf{g} Performance impact of different self-supervised learning strategies on classification and segmentation tasks. All results are scaled and normalized relative to UltraFedFM. Specific quantitative results are available in Supplementary Table 6. 
    }
    \label{fig:discussion}
\end{figure}

{Fig. \ref{fig:discussion}e-f} presents the scaling efficiency of UltraFedFM during pre-training, including data scaling ({Fig. \ref{fig:discussion}e}) and model size scaling ({Fig. \ref{fig:discussion}f}). Data scaling experiments were conducted using different proportions of pre-training data, while model size scaling involved pre-training with encoder architectures of varying parameter sizes (ViT-Base, ViT-Large, and ViT-Huge).
In the data scaling experiments, we randomly sampled 10\%, 20\%, 50\%, and 100\% of pre-training data from each client and evaluated performance on eight downstream classification tasks and five downstream segmentation tasks. Overall, increasing the amount of pre-training data improved model performance, consistent with the data scaling principles in self-supervised learning (SSL). However, the growth trends varied slightly across different data modalities. Notably, segmentation tasks exhibited more pronounced performance gains, indicating that high-dimensional pixel-level prediction tasks are more sensitive to pre-trained feature learning.
In the model size scaling experiments, we used three ViT variants as encoders to evaluate the impact of increasing the number of trainable parameters during pre-training on classification and segmentation tasks. For classification tasks, larger models generally yielded better performance across most modalities. Specifically, performance gains were more significant for challenging tasks (those with lower AUROC for ViT-Base), while simpler tasks reached a performance plateau, with ViT-Huge potentially introducing noise and overfitting risks. For segmentation tasks, increasing model size consistently improved performance, demonstrating that segmentation tasks demand larger model capacity and that model size scaling is particularly beneficial for addressing more challenging tasks.

{
\subsection*{Ablation studies validate the effectiveness of proposed strategies}
To thoroughly evaluate the contribution of each proposed module, we implemented eight ablated variants of UltraFedFM by replacing individual components and evaluated their performance on both classification and segmentation tasks, as shown in Fig. \ref{fig:discussion}d. Compared to the baseline, incorporating only SMAT (w/SMAT) yielded an increase of 5.5\% AUROC and 3.0\% DSC, while including only MIC (w/MIC) resulted in a more substantial improvement of 6.9\% AUROC and 3.9\% DSC. In contrast, removing specific components (i.e., w/o SMAT, w/o MIC, w/o TGM) degrades performance compared to the full UltraFedFM model. Notably, excluding MIC led to the highest decrease of 1.3\% AUROC and 3.4\% DSC, appearing to exert a particularly notable influence on the model's efficacy as evidenced by the relatively high performance when it is singularly included and the performance retention when other components are removed.
}
To validate the effectiveness of our modified masked autoencoder

\noindent (MAE) strategy in UltraFedFM, we compare it with several baseline self-supervised learning (SSL) strategies, including vanilla MAE \cite{MAE}, SimCLR \cite{SimCLR}, SwAV \cite{SwAV}, DINO \cite{DINO}, and MoCo \cite{MoCo}. {Fig. \ref{fig:discussion}g} shows that UltraFedFM with the modified MAE significantly outperforms all other baselines ($p < 0.001$) in both disease diagnosis and lesion segmentation tasks. Specifically, UltraFedFM achieves the highest average AUROC of 0.926 across eight classification tasks and an average DSC score of 0.878 across four segmentation tasks. In contrast, the vanilla MAE achieves the second-best performance, with an average AUROC of 0.884 and an average DSC score of 0.839. These results suggest that MAE-based approaches are more effective for ultrasound imaging than contrastive learning-based methods. This success may be attributed to MAE's ability to learn robust feature representations in images where structures can vary significantly across patients or organs. Clinically, this translates to more accurate diagnostic predictions, especially for complex cases involving subtle lesions or challenging anatomical regions.

{
\subsection*{Qualitative model analysis and visualization}}
{
UltraFedFM's performance on downstream tasks depends on the representations learned during training. To investigate how these representations support downstream decisions, we qualitatively analyzed the internal mechanisms of the pre-text task of UltraFedFM during pre-training and how UltraFedFM made task-specific decisions on downstream tasks.}

{
% pretraining
During pre-training, the pre-text task enables the model to learn ultrasound-specific context across various ultrasound imaging modalities. As shown in {Fig. \ref{fig:visualization}a}, UltraFedFM accurately reconstructs images even when large portions are masked while preserving anatomical textures and lesion structures. This qualitative reconstruction behavior suggests the model learns context-aware features that reflect tissue and lesion morphology rather than merely memorizing low-level noise patterns.
Fig. \ref{fig:visualization}b shows the results of scanning mode-aware transformation (SMAT) used in pre-training. Balancing scan modes reduces the tendency of UltraFedFM to overfit to a single probe configuration and promotes robustness across acquisition settings.}

% fine-tune: segmentation
For downstream lesion segmentation tasks, {Fig. \ref{fig:visualization}c} and {Fig. \ref{fig:visualization}d} visualize UltraFedFM's precise localization of salient lesion areas and target boundaries. Clinically, the model's ability to focus on salient areas while excluding irrelevant background interference enhances its accuracy in detecting complex lesion structures, which is essential for diagnosing diseases with subtle or overlapping symptoms.
% fine-tune: classification
Supplementary Fig. 1 illustrates the embedding feature space of different classes in the fine-tuned model. UltraFedFM demonstrates superior class discrimination, with different classes clearly separated in high-dimensional space, resulting in more precise classification boundaries. This ability is crucial in clinical applications where precise differentiation between pathological and non-pathological tissue can impact treatment decisions. In contrast, the baseline supervised model exhibits a weaker differentiation and less distinct classification results. Supplementary Fig. 2a further shows how UltraFedFM effectively recognizes specific patterns and targets via the attention mechanism. For example, in pancreas, liver, breast, and gallbladder imaging, the model focuses on the center and surrounding areas of tumor lesions. In colorectal and lung imaging, it targets high-density textured regions while ignoring irrelevant hollow regions such as intestines and alveoli. For fetal ultrasound imaging, UltraFedFM focuses on the solid parts of the fetus and excludes irrelevant regions such as the uterus and muscles. In addition, we visualized the evolution of the attention map during pre-training (see Supplementary Fig. 2b). As pre-training progresses, the local model increasingly focuses on meaningful regions, thereby enhancing the effectiveness of the global model.

\section*{Discussion}
\label{discussion}
With the growing demand for public health solutions, there is an urgent need to develop AI-based foundation models for wide application to real-world clinical scenarios. In this work, targeting the most widely used ultrasound data, we are the first to propose a comprehensive privacy-preserving ultrasound foundation model (USFM) using federated learning, namely UltraFedFM.
By eliminating privacy concerns through decentralized pre-training, UltraFedFM leverages large-scale global datasets, enhancing its generalization capabilities. Regarding the above extensive experimental results, UltraFedFM demonstrates excellent performance, favorable generalization and robustness, and good adaptability to fine-tuning data. Specifically, it can handle different clinical tasks, such as diagnosing diseases, segmenting regions of interest (i.e., pathological tissues or organs), and analyzing spatial relationships of fetal organs, making it versatile for a wide range of medical applications.
Moreover, UltraFedFM can be fine-tuned in a modality-agnostic manner to enable a single decoder to diagnose multiple diseases present in different modalities. Even with limited fine-tuning data, it consistently outperforms other baseline methods in both accuracy and stability. Across various ultrasound modalities and clinical tasks, UltraFedFM performs with judgment capabilities comparable to human clinicians.

% \textbf{The relations of UltraFedFM with previous achievements}. 
Ultrasound imaging, a widely used clinical diagnostic tool, is renowned for its convenience and accuracy. 
{Previous research in ultrasound diagnostics primarily focused on deep learning models trained on specific ultrasound modalities, targeting the diagnosis or segmentation of disease types within fixed imaging contexts. For example, Antropova et al.\cite{antropova2017deep} developed a method that utilized pre-trained CNNs to extract and aggregate features, which were then combined with hand-crafted features from CADx for breast cancer diagnosis. Similarly, Basu et al.\cite{gbcu} investigated multi-scale and second-order pooling architectures to address false textures in gallbladder ultrasound, achieving precise localization and detection of malignant gallbladder tumors. Objective assessments and specialized segmentation strategies have been proposed by Yadav et al. \cite{yadav2023objective, yadav2024deep, yadav2022despeckling, yadav2024machine} for thyroid ultrasound, highlighting the importance of robust segmentation and preprocessing in downstream classification. Comparative studies \cite{yadav2022despeckling, dass2020image, virmani2019assessment} on despeckling filters and image-quality metrics show that noise suppression and texture preservation strongly affect diagnostic}

\begin{figure}[H]
    \centering
    \includegraphics[width=\linewidth]{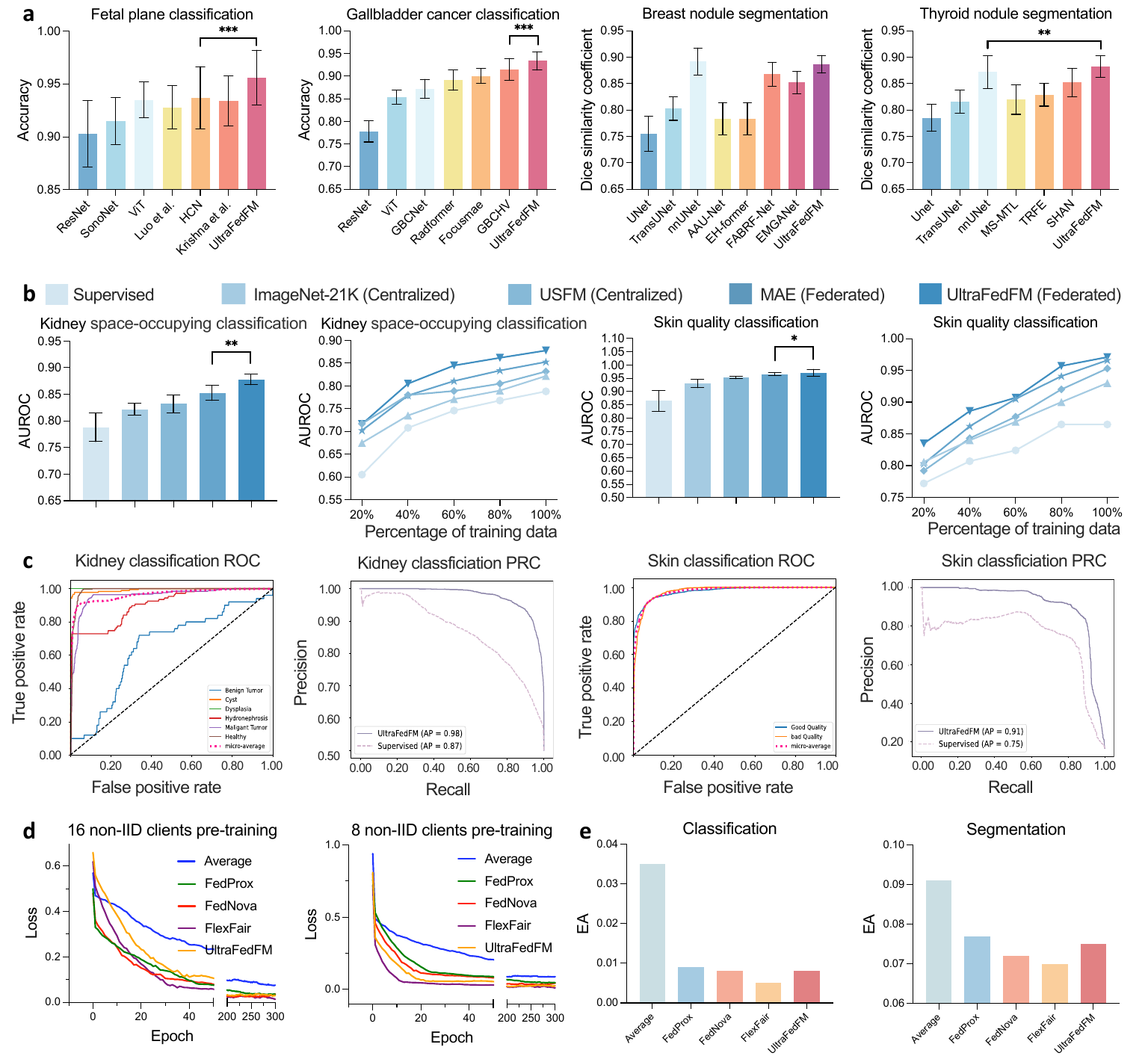}
    \caption{
    \textbf{a} The quantitative comparison between UltraFedFM and state-of-the-art ultrasound task-specific models across four typical ultrasound tasks. \textbf{b} Generalization performance evaluation on out-of-distribution organ (i.e., kidney) and modality (i.e., high-frequency ultrasound). \textbf{c} The receiver operating characteristic curve (ROC) and precision-recall curve (PRC) on a new organ and new ultrasound imaging modality. \textbf{d-e} The quantitative analysis of UltraFedFM with state-of-the-art federated learning methods in terms of pre-training convergence (d) and prediction fairness (e).
    }
    \label{fig:generalization}
\end{figure}

\noindent {performance. Several recent works also proposed deep-learning-based CAD systems with novel attention mechanisms for systematic disease diagnosis in ultrasound images. Jiang et al. \cite{ASTR} introduced a sparse computation and temporal fusion architecture designed for the accurate and real-time segmentation of colorectal cancer lesions. Yan et al. \cite{yan2025development} constructed and validated a deep learning-based radiomics fusion model, enabling accurate identification of bone erosions in rheumatoid arthritis in musculoskeletal ultrasound.} These studies have significantly advanced the application of AI in various ultrasound modalities, leading to improvements in automatic ultrasound image analysis, including lesion segmentation, disease diagnosis, and treatment planning. These successes are primarily attributed to the synergistic effects of data, models, and algorithms, specifically the collection and thorough annotation of datasets for specific organs or diseases, as well as specially designed network structures and training methods. However, a critical bottleneck limiting the advancement of medical imaging algorithms is the limited availability of fully annotated medical data. Medical image annotation demands the expertise of trained physicians,  

\begin{figure}[H]
    \centering
    \includegraphics[width=\linewidth]{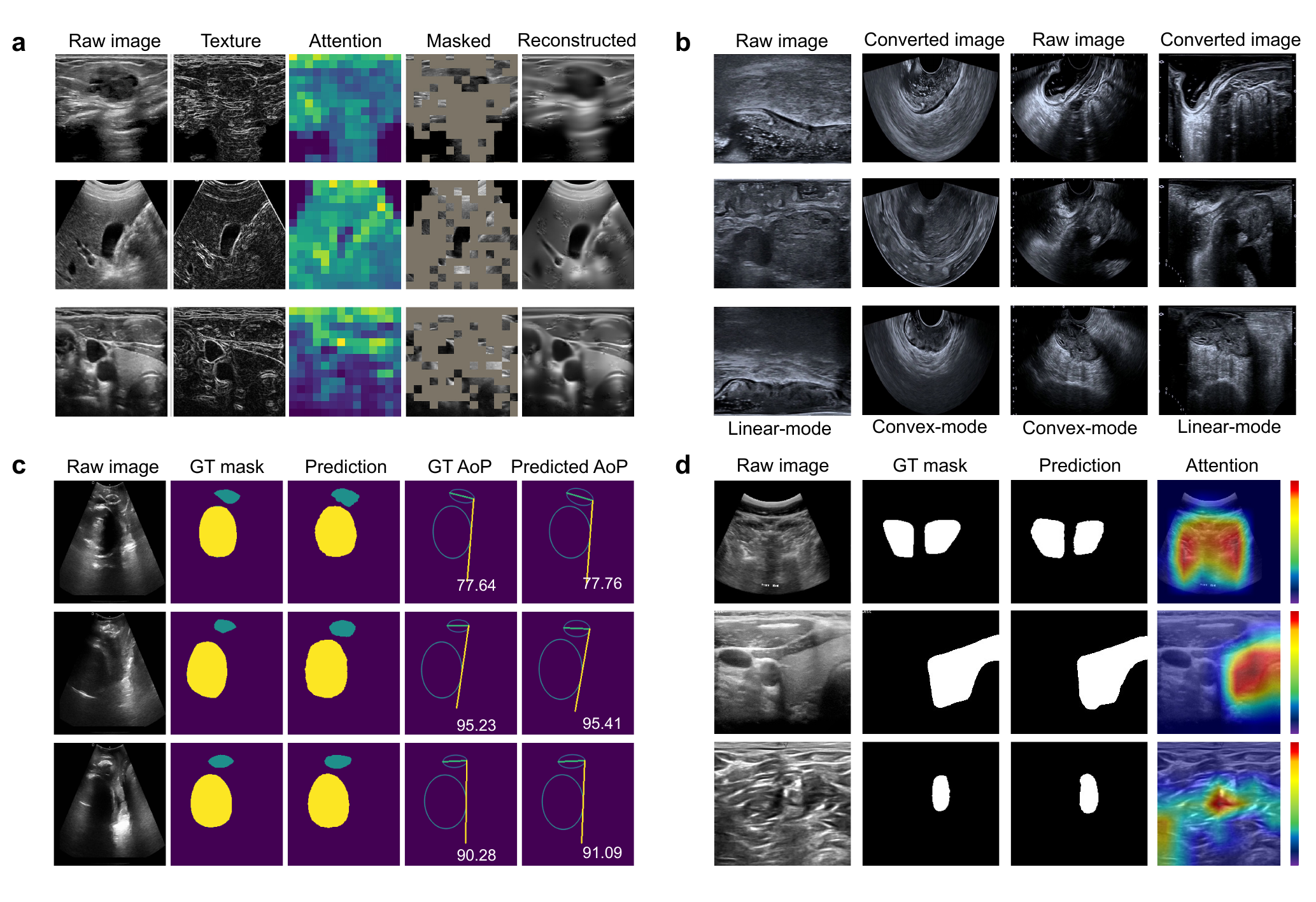}
    \caption{
    \textbf{a} The reconstructed ultrasound images from the pre-trained model, where the masked regions are selected based on texture information. 
    \textbf{b} To increase the richness and balance of features, images captured in linear-array mode and convex-array mode are transformed into each other. \textbf{c} Visualization of multi-class organ segmentation and the prediction of the angle of progression (AoP). \textbf{d} Visualization of binary lesion segmentation. Heatmaps highlight the attention areas of the features extracted from the pre-trained encoder. The closer the color is to red, the more the model pays attention to the area.}
    \label{fig:visualization}
\end{figure}

\noindent and the segmentation tasks in particular require substantial time and effort. Traditional methods were often trained on only hundreds or thousands of samples, significantly impacting the models' stability and generalizability in real-world applications. Meanwhile, with the evolution of ultrasound imaging technology, its application to an increasing number of organs and diseases presents challenges for models trained on single-organ or single-disease data, making it difficult to meet expanding clinical demands. There is a growing interest in developing a label-efficient ultrasound model that can be generalized across various tasks and organs, enabling rapid adaptation and deployment in clinical practice. 

This led to the introduction of the foundation model (FM) based on self-supervised learning (SSL), which is capable of learning universal features independent of organs and diseases from unlabeled data. Prior to UltraFedFM, several studies explored FMs in medical imaging~\cite{RetFound, VisionFM, EndoFM, CTFM, BioFM}, covering various modalities such as ophthalmic images, endoscopy, and CT scans. In the field of ultrasound imaging, Christensen et al.~\cite{EchoFM} proposed an FM specifically for cardiac ultrasound, pre-trained on over one million echocardiogram videos for diagnosing various heart diseases. Jiao et al.~\cite{USFM} compiled and organized over two million ultrasound images across 12 different categories to establish a general USFM. However, these foundation models either focus on developing and applying to a single ultrasound modality, which is limited in actual clinical deployment or require centralized collection and processing of multi-center large-scale data. On the one hand, this approach requires expensive servers to store and process data, and on the other hand, the circulation of data inevitably involves the disclosure of patient information, especially rare disease data, which hinders the development of universal models.

% \textbf{The advantage of UltraFedFM}. 
UltraFedFM confirms that distributed pre-trained foundation models can match centralized ones by appropriate design. While UltraFedFM is not the first FM developed for ultrasound imaging, it is the first to integrate privacy protection during the model development process. Previous methods highlighted that, in most cases, private data held by different institutions is not shared, and public data must be anonymized to protect patient privacy, making it challenging to utilize a vast amount of available medical data. 
{In this study, UltraFedFM breaks the privacy obstacles by leveraging federated learning that can use large inaccessible private data to train model distributively across sites without sharing sensitive information. UltraFedFM was pre-trained on over 1 million ultrasound images from 16 independent institutions worldwide, covering 19 different organs and 10 ultrasound modalities, which is $58.3\%$ more in organ coverage and $66.7\%$ more in ultrasound modalities compared with centralized baseline 3M-US, and therefore captures a substantially wide range of acquisition devices, operators and clinical scenarios. To address ultrasound-specific pre-training challenges, UltraFedFM developed the ultrasound image masking strategies, which explicitly accounts for probe-dependent texture characteristics and reduces failure modes observed when using generic MAE on ultrasound data. More importantly, the federated architecture also confers practical extensibility, where new private datasets can be incorporated incrementally to update the foundation model, so the utility of the model can increase overtime as more institutions join. Taken together, through careful selection of pre-training image quantity and the design of pre-training algorithms specific to ultrasound imaging, it has been demonstrated for the first time that a distributed pre-training foundation model can achieve comparable performance to centralized foundation models in overall performance across multiple downstream tasks. In ultrasound disease diagnosis tasks, UltraFedFM achieved an average AUROC of 0.927 across eight organs, significantly ($p<0.001$) surpassing USFM's 0.894 AUROC. In ultrasound lesion segmentation tasks, UltraFedFM achieved an average DSC score of 0.876, significantly ($p<0.001$) exceeding USFM's 0.858.}

% {\textbf{UltraFedFM's federated learning enables privacy-compliant, collaborative medical AI model development.}}
The advantages of UltraFedFM go beyond its performance metrics. The federated learning framework it employs provides unique benefits that address long-standing challenges in medical AI. First, it enables the model to be continuously updatable, allowing further training and refinement on private datasets held by individual institutions or clients without exposing sensitive data. This ensures compliance with privacy regulations, such as GDPR and HIPAA, while maintaining the model's adaptability to evolving clinical needs. Additionally, federated learning facilitates participation from small-scale data contributors with uncommon organs or rare modalities (e.g., uterus, testis, contrast-enhanced ultrasound, high-frequency ultrasound). This fosters a collaborative ecosystem where diverse and distributed medical data can be leveraged to develop a comprehensive and generalizable model. These capabilities establish UltraFedFM as a solution capable of bridging the gap between data privacy and large-scale model training.
The security of UltraFedFM during the pre-training process and its accuracy across various ultrasound imaging applications further solidify its potential for clinical translation. Previously, only large institutions with efficient data management workflows could develop foundation models from vast private medical datasets. This study demonstrates that federated learning enables the global medical community to collectively pre-train robust and generalizable foundation models. 
By addressing the dual imperatives of privacy preservation and model performance, UltraFedFM marks a paradigm shift in medical AI. Its ability to adapt to new data while protecting patient privacy sets a benchmark for the development of privacy-preserving AI in healthcare. Through its innovative use of federated learning, UltraFedFM demonstrates how global collaboration can unlock the potential of distributed medical data, paving the way for advancements in ultrasound AI and broader medical applications.

% {
% \textbf{Challenges of non-independent and identically distributed data.}
% The core innovation of this study lies in using a federated learning framework to simulate the distributed data distribution in real clinical environments. 
{The core innovation of this study is the application and engineering of a large-scale, ultrasound-specific federated pre-training framework to simulate the distributed data distribution in real clinical environments.}
During the federated pre-training stage, the ultrasound image data of each participating institution showed significant differences in both sample size and modality type, leading to the typical non-independent and identically distributed (non-IID) characteristics of the data. The sensitive attribute distribution shown in Fig. \ref{fig:dataset}c further confirms the systematic bias existing in the dataset, which poses a substantial challenge for cross-institutional model aggregation. To systematically evaluate the model's adaptability to non-independent and identically distributed (non-IID) data, we designed two sets of experiments focusing on convergence and fairness, respectively. For the convergence validation, we implemented two testing scenarios: the first using original non-IID training data from 16 different institutional clients for pre-training, while the second employed randomly sampled data from 8 institutions. UltraFedFM was compared against four federated learning approaches: Average (the baseline method without volume-based weighting strategy), along with state-of-the-art methods specifically designed for non-IID scenarios, including FedProx \cite{li2020fedprox}, FedNova \cite{wang2020fednova}, and FlexFair \cite{xing2025flexfair}. Experimental results in Fig. \ref{fig:generalization}d demonstrated that UltraFedFM significantly outperformed the Average baseline in both scenarios. Specifically, in the 8-client experiment, FlexFair achieved the fastest convergence speed, followed by UltraFedFM, whereas in the more challenging 16-client experiment, UltraFedFM showed a comparable convergence speed to FedProx, both substantially surpassing the simple averaging method. These findings indicate that the uniform weighting approach of simple averaging causes deviation of the global objective function from its optimal direction, thereby reducing the convergence rate, while simultaneously confirming UltraFedFM's robustness in handling non-IID data. For fairness evaluation, we constructed simulated non-IID experimental environments. The classification task utilized three breast ultrasound datasets (BUS \cite{busi}, BUS-BRA \cite{gomez2024bus-bra}, and BUS-UCLM \cite{vallez2025bus-uclm}) as clients, while the segmentation task employed three thyroid ultrasound datasets (DDTI \cite{ddti}, TG3k \cite{tg3k}, and TN3k \cite{tn3k-1, tn3k-2}). Each client dataset was randomly split into 80\% training and 20\% testing sets. Equal Accuracy (EA) \cite{xing2025flexfair} served as the fairness metric, measuring maximum prediction accuracy disparities across different groups (e.g., hospitals or age cohorts). Results in Fig. \ref{fig:generalization}e showed FlexFair achieved optimal EA fairness performance in both tasks. UltraFedFM ranked second in classification fairness and performed comparably to FedProx in segmentation. Notably, the simple averaging approach performed poorly in both experiments, conclusively demonstrating the necessity of weighted aggregation strategies for non-IID data. Weighting by client data volume effectively prevents information dilution from large-data clients and substantially mitigates negative impacts from extreme data distributions.

% \textbf{Current limitations of UltraFedFM}. 
While systematically evaluating the advantages of UltraFedFM in ultrasound imaging analysis, this study still has several limitations and unresolved challenges. 
Firstly, although currently we use federated learning to simulate distributed privacy-preserving training to avoid data information leakage between clients. However, the deployment is not in a fully decentralized clinical environment, and it cannot fully replicate the challenges, such as heterogeneous device differences, network communication delays, and dynamic client participation in a real multi-center scenario. This limitation further highlights the necessity of establishing an inter-institutional joint research network and promoting real distributed training.
Secondly, although UltraFedFM has established a comprehensive ultrasound imaging benchmark encompassing 10 ultrasound modalities and 19 human organs, demonstrating outstanding performance in relevant tasks, its data diversity remains limited. Compared to the broader scope of clinical data, the benchmark lacks coverage in certain critical ultrasound imaging domains (e.g., rheumatoid arthritis ultrasound, ocular ultrasound, etc.), primarily due to the scarcity of publicly available datasets. This limitation hinders the full validation of UltraFedFM's generalization capability in these scenarios. Furthermore, the current version of clinical evaluation relies on a relatively small sample size and a limited number of participating physicians, making it difficult to comprehensively reflect the diversity and complexity of real-world clinical practice. Moving forward, we plan to collaborate extensively with clinical institutions to collect multicenter clinical samples covering mainstream ultrasound modalities and rare diseases, while inviting clinical experts from diverse institutions to participate in larger-scale double-blind evaluations. These efforts aim to ensure seamless integration of the model into clinical ultrasound workflows and strict compliance with clinical standards and practical requirements.
% Thirdly, while UltraFedFM has demonstrated effectiveness in handling non-IID data, the increasing scale of healthcare resources and growing number of participating institutions will inevitably lead to greater demands for client-side communication in future applications. The current simplified federated architecture may prove insufficient to address the complexities and scale requirements of large-scale non-IID pre-training scenarios. Future research efforts should therefore prioritize the development of more sophisticated algorithms capable of operating effectively in real-world distributed environments with heterogeneous data distributions.
{Thirdly, while our simulated non-IID experiments indicate some robustness of the current volume-weighted federated learning setup, its effectiveness may decline as institutions and heterogeneity grow. Future research efforts should therefore explore targeted extensions with specific fairness designs, such as fairness-aware aggregation to balance overall per-client equity, and scalable federation to improve scalability in real-world deployments.}
Lastly, although UltraFedFM effectively addresses critical clinical diagnostic and segmentation tasks and has been validated across 12 different types of organs and diseases, it does not utilize the vast amount of textual diagnostic reports available in ultrasound examinations. Integrating multimodal features from both text and images could further improve the foundational model's accuracy in various clinical tasks and enable the development of additional clinically useful tasks like question-answering and diagnostic report generation.

% \textbf{The significance and application of UltraFedFM}. 
In conclusion, this study provided a robust and reliable framework for developing comprehensive USFMs. It has been demonstrated that high-performance and stable models can be pre-trained without risking privacy leakage. This breakthrough can significantly advance the development of USFMs and potentially lead to the emergence of more powerful general-purpose medical AI. UltraFedFM has already shown excellent performance in various ultrasound clinical tasks and holds promise for further expansion. It has the potential to replace traditional ultrasound AI diagnostic models and play a crucial role in clinical decision support. The theoretical contributions of this research lie in validating the efficacy of federated learning for pre-training medical FMs, inspiring further academic advancements in the field. Practically, the implementation of UltraFedFM can enhance diagnostic accuracy, streamline clinical workflows, and improve patient outcomes. Additionally, these findings offer valuable insights for policy-making, particularly in the areas of data privacy and the integration of AI in healthcare. By ensuring data privacy and leveraging federated learning, the collective power of global medical data can be harnessed to drive innovations in medical diagnostics and treatment planning, ultimately transforming the healthcare delivery landscape.

\section*{Methods}
\label{method}
% distribution pretrain dataset development 
\subsection*{Ultrasound dataset curation}
UltraFedFM aims to perform universal ultrasound tasks for clinical applications. A crucial aspect of constructing such a model is the adaptability to diverse ultrasound imaging modalities and pathological conditions. To address this issue, we curated a large-scale pre-training dataset consisting of 1,015,754 unlabeled ultrasound images. In this dataset, 782,513 images were publicly available from multiple worldwide hospitals, while the remaining images were our privately owned ultrasound data. Our dataset covers a wide range of clinical ultrasound scenarios and modalities, including common abdominal, heart, fetal, superficial, musculoskeletal, and transvaginal ultrasound, as well as emerging techniques such as lung, endorectal, endoscopic, high-frequency, and contrast-enhanced ultrasound ({Fig. \ref{fig:dataset}a}). Moreover, to realize federated pre-training, we split and arranged all datasets into 16 clients according to different hospitals. Supplementary Table 1 presents the data distributions of all clients. The multi-organ and multi-modality categories enable UltraFedFM to adapt to a wide range of clinical tasks.

{
% 数据集包含广泛的真实敏感属性，其中一些属性存在偏差。
The dataset contains a wide range of real sensitive attributes, and some of these attributes are biased. As illustrated in Fig. \ref{fig:dataset}c. Specifically, at the national level, data from Chinese medical institutions accounts for the largest proportion at 57.30\%, primarily because most private data comes from collaborative Chinese institutions. Followed by the United States at 15.94\% and Germany at 10.14\%, with all other countries comprising less than 20\%. This imbalance may lead UltraFedFM to perform better on Chinese patient cases while introducing bias against underrepresented regions. In terms of geographical population distribution, the majority of cases come from Asia, followed by Europe and the Americas, reflecting imbalances in global healthcare resources. At the gender level, although the known gender ratio appears balanced, the large portion of samples with unspecified gender raises concerns about hidden gender bias. In terms of ultrasound modalities, mainstream modalities (such as abdominal, superficial, etc.) are relatively evenly distributed, while other modalities account for a smaller proportion due to their lower usage frequency. The data also primarily uses conventional linear scanning modes. Regarding the image quality of downstream fine-tuning data, nearly 50\% of the data has varying degrees of quality issues, which helps us evaluate whether the model exhibits bias with low-quality data. In future work, in addition to expanding the geographical coverage and modality range of the data to minimize information bias introduced by the data, we also plan to explore the effectiveness of fairness-aware federated learning in improving dataset bias.
}

To further verify the practicality of UltraFedFM on clinically relevant tasks, we curated 15 well-annotated ultrasound datasets for validation. Two common clinical tasks were tested. The first task is ultrasound image diagnosis, which requires the FM to make accurate category judgments based on the organ and lesion information, ranging from two-class cancer recognition to multi-class disease diagnosis ({Fig. \ref{fig:dataset}b top}). For this task, we utilized 7 publicly available datasets and 3 internal datasets for validation, which included a total of 10 organ categories and 8 ultrasound modalities. The second task is ultrasound image segmentation, which requires the FM to identify key organ/lesion areas and predict boundaries. For this task, we collected 5 public datasets for validation, which included a total of 5 organ categories and 4 ultrasound modalities. Note that small targets are more challenging for the model's prediction ability. In the validation datasets, 64.5\% of the images contained targets that are less than 1/10 of the total image area ({Fig. \ref{fig:dataset}b bottom}).

The overall ultrasound pre-training dataset and validation dataset breakdown are presented in Supplementary Table 1 and Supplementary 2.

{
\subsection*{Clinician cohort}
To truly evaluate the reliability of UltraFedFM as an auxiliary tool in clinical scenarios, we invited multiple clinicians to participate in the evaluation.
All participating clinicians hold a physician qualification certificate and a physician's practice certificate issued by the National Health Commission of China, and have completed subspecialty fellowship training in abdominal or musculoskeletal ultrasound. In addition, mid-level clinicians (with 4-8 years of clinical experience) hold a certificate of qualification for intermediate professional and technical positions (CQIPTP), and their professional titles are attending physicians. Expert-level doctors (with $10$ years of clinical experience) hold a certificate of qualification for senior professional and technical positions (CQSPTP), and their professional titles are chief physicians or associate chief physicians. Additionally, expert-level clinicians handle an average of $7,000 \pm 200$ ultrasound clinical cases per year (with an average of 240 working days), whereas mid-level clinicians (4–8 years, n=2) average $4,500 \pm 300$ clinical cases per year.
}

\subsection*{Federated pre-training framework}
This work aims to collaborate with multiple clients to jointly train a robust FM without sharing their privacy-sensitive ultrasound data. 
% Therefore, we employ a federated pre-training architecture consisting of $K$ local clients and one server. 
{To simulate the real clinical decentralized setting, we partitioned the pre-training dataset in $K$ ``virtual clients'' to mimic independent institutions. Our simulation ensures raw data never leaves clients and assumes encrypted communication channels.}
Each client $k \in \{1, \cdots, K\}$ possesses a local dataset $\mathcal{D}_k$ with $n_k$ data samples. The objective is to learn a global model, consisting of a global encoder $\mathtt{E}_g$ and a global decoder $\mathtt{D}_g$ towards minimizing the global loss function, which can be expressed as
\begin{equation}
    L(\omega) =  \frac{1}{n} \sum_{k=1}^K n_k L_k(\omega),
\end{equation}
where $\omega$ is the overall model parameters of the global encoder $\mathtt{E}_g$ and global decoder $\mathtt{D}_g$, $n= \sum_{k=1}^K n_k$ is the total number of data of all clients, and $L_k(\omega)$ is the local loss function of client $k$ that measures the local empirical loss on its local dataset $\mathcal{D}_k$. Then, the pre-training stage can be described as follows.
\begin{enumerate} \itemsep -1.5pt
    \item In the $t$-th communication round, the server broadcasts the global model $\omega^t$ to all clients.
    \item Each client $k$ takes $E$ steps of gradient descent to update the local model based the received global model $\omega^t$, as given by
    \begin{equation}
    \omega_k^{t+1} = \omega_k^t - \eta \nabla L_k(\omega^t),
    \end{equation}
    where $\omega_k$ denotes the local model parameter of client $k$ and $\eta$ is the learning rate.
    \item Each client uploads its local model parameter $\omega_k^{t+1}$ to the server.
    \item The server aggregates the local models from all clients and updates the global model by
    \begin{equation}
    \omega^{t+1} = \frac{1}{n} \sum_{k=1}^K n_k \omega_k^{t+1}.
    \end{equation}
\end{enumerate}
The combination of the four steps is referred to as one communication round. The pre-training process terminates once it reaches a pre-defined number of communication rounds $T$.

After completing pre-training, the pre-trained global encoder was saved while the global decoder was discarded.
In the fine-tuning stage, the global encoder generated high-level features from the ultrasound images. 
A multi-layer perceptron (MLP) takes these features as input and outputs the probabilities for disease categories. 
The category with the highest probability was defined as the classification result.
The fine-tuning objective was to produce classification results that match the ground-truth labels.
After each epoch, the model was evaluated on the validation set. The model weights with the highest accuracy on the validation set were saved as checkpoints for internal and external evaluations.

\subsection*{Local model architecture}
As illustrated in Supplementary Fig. 3, we employed a masked autoencoder (MAE) as the local model of each client $k$, which consists of an encoder $\mathtt{E}_k$ and decoder $\mathtt{D}_k$. 

For each image $I \in \mathbb{R}^{H \times W \times C}$ sampled from the local dataset, the corresponding local model undergoes pre-training through an ultrasound masked image modeling (UltraMIM) process, including an ultrasound image masking (UIM) stage and a reconstruction stage, which are respectively defined by
\begin{align}
        \mathcal{P} &= \mathtt{UIM}_k(I),\\
        I^{recon} &= \mathtt{D}_k( \mathtt{E}_k (\mathcal{P})),
\end{align}
where $\mathcal{P} = \{\mathbf{p}^1, \cdots, \mathbf{p}^L \}$ is the patch set, $\mathbf{p}^{\ell} \in \mathbb{R}^{N}$ represents the $\ell$-th patch, $L$ denotes the total number of patches, and $\mathtt{UIM}$ denotes the ultrasound image masking operation. In particular, there exist two types of patches in the $\mathcal{P}$, i.e., masked patches $\mathcal{P}_m$ and visible patches $\mathcal{P}_v$.

We utilized Vision Transformer (ViT) as the encoder $\mathtt{E}_k$ and applied it to a sequence of unmasked image patches. Specifically, the encoder compressed the input visible patches $\mathcal{P}_v$ into the latent representation, denoted by $\mathcal{P}_v^{\prime}$. 
%by $\mathcal{R}=\mathtt{E}_k(\mathcal{P}_v)$. 
The latent representation captures the essential features of the input images, allowing the model to learn meaningful patterns regardless of masking. 
%To deal with the masked regions, we then inserted mask tokens into the positions corresponding to the masked patches as $z^{\prime}=[z,~\text{mask tokens}]$. These tokens serve as placeholders, enabling the model to learn to reconstruct the missing information during the decoding process. Thereafter, the combined latent representation $z'$
Then, we concatenated $\mathcal{P}_v^{\prime}$ and $\mathcal{P}_m$ to obtain the overall patch set, defined as $\mathcal{P}^{all} = \mathtt{Concat}(\mathcal{P}^{\prime}_v, \mathcal{P}_m)$. After that, the overall patch set $\mathcal{P}^{all}$ was passed through the decoder $\mathtt{D}_k$, which is a lightweight ViT that aims to reconstruct the input image by predicting the pixel values of the masked patches. Finally, the output of the decoder was reshaped to obtain a reconstructed image. We note that the reconstruction process enables the model to learn the underlying structure and patterns of the ultrasound images. Without loss of generality, we adopted the mean square error (MSE) as the local loss function to measure the discrepancy between the original image and the reconstructed image, which is defined as
\begin{equation}
    L_k(\omega)  = \sum_{1 \leq i \leq N_m}\frac{1}{N_m}\left(x^i-\hat{x}^i\right)^2,
\end{equation}
where $N_m$ is the number of masked patches, while $x$ and $\hat{x}$ are the ground-truth and predicted pixel values of each masked patch.

% adaptive mask image modeling 
\subsection*{Ultrasound image masking}
%-----------------------------------------------------------%
In clinical scenarios, different medical institutions may use different probes and equipment to collect ultrasound images. As a result, there exists significant variability in the imaged organs and lesions. To enhance the generalization ability of UltraFedFM across diverse clinical scenarios, we proposed the UIM composed of three modules, including scanning mode-aware transformation (SMAT), mixed image corruption (MIC), and texture-guided masking (TGM), which adaptively adjust the pre-training process based on the characteristics of the local dataset. Specifically, the working mechanism of UIM at client $k$ can be expressed as
\begin{align}
        \mathcal{D}_k^{trans} &= \mathtt{SMAT}(\mathcal{D}_k), \\
        \mathcal{D}_k^{total} & = \mathtt{Concat}(\mathcal{D}^{trans}_k, \mathcal{D}_k), \\
        I^{corru} &= \mathtt{MIC}(I, p), \\
       \mathcal{P} &= \mathtt{TGM}(I^{corru}),
\end{align}
where $\mathcal{D}_k^{trans}$ is the transformed dataset, $\mathcal{D}_k^{total}$ is the overall dataset for pretraining, and $I^{corru}$ is the corrupted input image.

\noindent \textbf{Scanning Mode-Aware Transformation.}
%-----------------------------------------------------------%
To ensure the generalization ability of UltraFedFM to images acquired under different scanning modes, we introduced a data augmentation method called Scanning Mode-Aware Transformation (SMAT). Typically, ultrasound images are collected using convex-array or linear-array probes, each with distinct geometrical properties. Therefore, we leveraged the coordinate mapping relationship between the two modes through Polar-Cartesian transformations \cite{polar2cartesian}.

Supplementary Fig. 4a shows the detailed process of SMAT. 
%For each image $I$ in the original dataset $D_k$, 
We first established the Polar coordinates for the convex-array mode image using the origin point $(r_o, \theta_o)$ set at the top center of the image and the x-axis at the top edge. Meanwhile, the Cartesian coordinates are set using $(x_o, y_o)$ as the origin point, the top edge as the x-axis, and the vertical central axis as the y-axis. 
In Cartesian coordinates, we denote $(x_1, y_1)$ as the point in the original image, $(x_2, y_2)$ as the point in the transformed image. In polar system,  $(r_1, \theta_1)$ is the point in the original image, $(r_2, \theta_2)$ is the point in the transformed image.
For the convex-array mode, we transform the Polar coordinate into the Cartesian coordinate.  
Specifically, we obtained the value and position of each pixel $(x_2,y_2)$ according to its corresponding point $(r_1,\theta_1)$ on the original image: 
\begin{align}
    r_1 &= \sqrt{x_1^2 + y_1^2}, \\
    \theta_1 &= \arctan\frac{y_1}{x_1}, \\
    I^{trans}[x_2, y_2] &= f(x_o+r_1\cos\theta_1,~y_o+r_1\sin\theta_1),    \label{equ:convex}
\end{align}   
where $I^{trans}$ denotes the transformed image, $f$ is a function for getting the information of the corresponding pixel points in the input image. 

For the linear-array mode, the transformed position $(r,\theta)$ of the convex-array mode image is calculated using the Cartesian-to-Polar transformation, which can be expressed as 
    \begin{align}
    r_2 &= \sqrt{x_1^2 + y_1^2}, \\
    \theta_2 &= \arctan \frac{y_1}{x_1}, \\
    I^{trans}[x_2, y_2] &= f(x_o+r_2\cos\theta_2,~ y_o+r_2\sin\theta_2).  
    \end{align}   

%Let $\mathcal{D}_k^{\text{A}}$ denote the augmented dataset of client $k$ that consists of all transformed images. 
After image transformation, we then concatenated the transformed dataset and the original dataset to obtain an enhanced pre-training dataset, i.e., $\mathcal{D}_k^{total}$. Note that the number of images in $\mathcal{D}_k^{total}$ is balanced across different scanning modes. Therefore, the risk of model over-fitting to any particular mode is significantly reduced.

\noindent \textbf{Mixed image corruption.}
%-----------------------------------------------------------%
To ensure that UltraFedFM is robust to both low-quality and high-quality images, we introduced an additional de-corruption branch in the pre-training process, as shown in Supplementary Fig. 4b. The image corruption operations were inspired by three common cases encountered in clinical practice, i.e., motion blur, low resolution, and random noise.
\begin{enumerate}\itemsep -1.5pt
    \item Motion blur arises from the swift movement of the ultrasound probe, leading to artifacts and image distortion. We simulated this effect by convolving the original image with a motion blur kernel $\mathbf{K}(d, \phi)$ that accounts for both the degree of blur $d$ and the angle of motion $\phi$.
    Specifically, we first constructed an identity matrix $\mathbf{U} \in \mathbb{R}^{d \times d}$.
    % where only diagonal elements are set to 1.
    Moreover, we defined a rotation matrix $\mathbf{R}(\phi)$ with respect to the motion angle $\phi$, as given by
    \begin{equation}
        \mathbf{R}(\phi) =
        \begin{bmatrix}
            cos(\phi) &-sin(\phi)\\
            sin(\phi) &cos(\phi) 
        \end{bmatrix}.
    \end{equation}
    Then, by applying the affine transformation, the motion blur kernel $\mathbf{K}(d, \phi)$ is derived as
    \begin{equation}
        \mathbf{K}(d, \phi) = \frac{1}{d} \cdot \mathbf{U} \cdot \mathbf{R}(\phi),
    \end{equation}
    where each element in $\mathbf{K}(d, \phi)$ was normalized by the degree of blur $d$ to ensure that the kernel maintains the same intensity as the original image.
    Finally, the motion-blurred image can be obtained by convolving the original image with the motion kernel, which is expressed as
    \begin{equation}
        I^{corru} = I \ast \mathbf{K}(d, \phi),
    \end{equation}
    where $\ast$ denotes the convolution operation.

\item Low resolution makes it hard to distinguish the critical parts of images. It can be achieved by a simple Gaussian blur operation. 
Gaussian blur involves convolving an image with a Gaussian kernel $G(\sigma)$, which is a two-dimensional Gaussian function with a standard deviation $\sigma$. 
Mathematically, the Gaussian blur operation can be expressed as
\begin{equation}
    \begin{aligned}
        I^{corru} &= \mathtt{Gaussian}(I, \sigma)\\
            &= I*G(u, v; \sigma) \\
            &= I*\frac{1}{2\pi\sigma^2}e^{-(u^2+v^2)/2\sigma^2},
    \end{aligned}
\end{equation}
where $u$ is the distance to the origin in the horizontal axis, $v$ is the distance to the origin in the vertical axis.

\item Random noise is the disturbance of pixel values, which is caused by the aging of components in old equipment. Random noise often destroys the appearance features of the target and affects the doctor's judgment. In this work, we simulated salt-and-pepper noise by the random occurrence of black and white pixels in an image. Given a grayscale image $I(x,y)$ where 
$(x,y)$ denotes the pixel coordinates. The corrupted image is given by
\begin{align}
    I^{corru}(x,y) = \left\{
        \begin{aligned}
         &0, &&\text{with~probability}~p_s,\\
        &255, &&\text{with~probability}~p_p,\\
        &I(x,y), &&\text{with~probability}~1-p_s-p_p,
        \end{aligned}
    \right.
\end{align}
where $p_s$ is the probability of a pixel being set to the minimum intensity value ("0" corresponds to "salt"), and $p_p$ is the probability of a pixel being set to the maximum intensity value ("255" corresponds to "pepper").
\end{enumerate}
 
The above three image corruption transformations can be combined to constitute a variety of composite transformations. We randomly selected one, two, or three operations from the triplet of [motion blur, gaussian blur, random noise] according to probability p to form a composite operation.
%(ensuring that there is at least one operation).

\noindent \textbf{Texture-guided masking.}
%-----------------------------------------------------------%
In ultrasound images, the edge prior reveals the sharpness of local regions and contains high anatomical information.  Therefore, we quantified the edge information of each image patch to measure the texture complexity. The process is illustrated in Supplementary Fig. 4c. Specifically, given an ultrasound image $I$ with the spatial size of $H \times W$, we computed the texture map $I^{texture} \in \mathbb{R}^{H \times W \times 1}$ based on the edge information of $I$ using a second-order Laplacian differential operator~\cite{li2023you}, as defined by
%, which has stable localization ability and sharpness~\cite{li2023you}.
%The texture extraction process is defined as:
\begin{equation}
   I^{texture} = \frac{\partial^2I}{\partial x^2} + \frac{\partial^2I}{\partial y^2},
\end{equation}
where $x$ and $y$ are the indices of the image $I$. 
Then we split $I^{texture}$ into a series of texture patches, denoted by a set $\mathcal{P}_t = \{\mathbf{p}_{t}^1, \cdots, \mathbf{p}_{t}^L\}$. Here, $\mathbf{p}_{t}^{\ell} \in\mathbb{R}^{h \times w \times 1}$ is the $\ell$-th texture patch with spatial size of $h \times w$. 
For each texture patch, the texture complexity score was calculated by summing the absolute values of all elements,
which is given by  
\begin{equation}
    \text{Score}^l_t = \sum^h_{i=1}\sum^w_{j=1} \Vert \mathbf{p}_{t}^l(i,j)\Vert,
\end{equation}
where $\text{Score}^l_t$ is the score of the $l$-th texture patch and $(i,j)$ denotes the position of the texture patch. 
%Here, we set the patch size as 16. 
By doing so, we can obtain the scores for all texture patches, denoted by $\left[\text{Score}^1_t, \text{Score}^2_t, ..., \text{Score}^L_t \right]$. Then, we generated the texture attention mask by concatenating the score of all texture patches, as given by
\begin{equation}
    \mathbf{A} = \mathtt{Concat}(\text{Score}^1_t, \text{Score}^2_t, ..., \text{Score}^L_t).
\end{equation}

Given the attention mask $\mathbf{A}\in\mathbb{R}^{L}$, the texture patches with higher weights are more likely to be the foreground critical objects and may contain more information than those with lower weights. Therefore, we raised the masking probabilities of texture patches with high weights and used the texture patches with low weights as the visible hints in the masked image modeling process. More precisely, we first sorted the values in $\mathbf{A}$ from largest to smallest. Then, we guided the token selection process to generate the masked patch set $\mathcal{P}$. The tokens with the top $M$ highest probabilities were discarded, while the remaining tokens were preserved as visible hints for masked image modeling. Here, we set $M=75\%$ to keep consistent with the original MAE settings.

% Contrastive SSL implementation
\subsection*{Self-supervised learning implementation}
%-----------------------------------------------------------%
For comparison purposes, we replaced the MAE \cite{MAE} in UltraFedFM, with other SSL methods, including SimCLR \cite{SimCLR}, SwAV \cite{SwAV}, DINO \cite{DINO}, and MoCo-v3 \cite{MoCo} to generate different pre-trained models. For each SSL method,
%training using each contrastive learning method, 
we followed the network architectures and hyperparameter settings recommended in the literature to achieve the optimal performance. First, we loaded the pre-trained weights on ImageNet-1k into the models. Subsequently, we trained the models using the ultrasound pre-training dataset with each SSL method to obtain the pre-trained models. Following the same process for UltraFedFM, we transferred the MAEs to downstream disease detection tasks and fine-tuned these pre-trained models.

\subsection*{Baseline methods implementation}
We compared the performance of UltraFedFM with 4 pre-trained comparison models: Supervised, ImageNet-21K (centralized), USFM (centralized), and MAE (federated). ``Supervise'' uses the supervised learning strategy with a randomly initialized ViT encoder. ``ImageNet-21K'' uses the transfer learning strategy, where the model is centralized pre-trained on ImageNet-21K (about 14 million natural images with classification labels) through self-supervised learning. ``USFM'' uses the Universal Ultrasound Foundation Model \cite{USFM} to perform centralized pre-training on the 3M-US dataset (about 2 million ultrasound images of 12 organs). MAE and UltraFedFM use the same ultrasound dataset for federated pre-training, but MAE uses the original masked image modeling algorithm as a control to observe the advantages of our newly designed modules. All methods are fine-tuned on the same downstream datasets using the same experimental settings until convergence.

% Evaluation metrics
\subsection*{Performance metrics}
To evaluate the performance of all disease diagnosis tasks, we utilized four widely employed metrics, including accuracy, F1-score, AUROC, and recall. 
Accuracy is a fundamental metric in classification tasks, which is defined as the ratio of correctly classified instances to the total number of instances. Mathematically, it is expressed as
\begin{equation}
    \text{Accuracy}  =\frac{TP+TN}{TP+TN+FP+FN},
\end{equation}
where $TP$, $TN$, $FP$, and $FN$ represent the true positives, true negatives, false positives, and false negatives, respectively. 

F1-score is often used to evaluate the performance of classification models, particularly in scenarios where the data distribution is imbalanced. It is defined as the harmonic mean of precision and recall, offering a balance between the two metrics. F1-score is mathematically expressed as
\begin{align}
    \text{F1-score} &= 2 \cdot \frac{ \text{Precision} \cdot \text{Recall}}{\text{Precision}+\text{Recall}}, \\
    \text{Precision} &= \frac{TP}{TP+FP}, \\
    \text{Recall} &= \frac{TP}{TP+FN}.
\end{align}

AUROC is utilized to assess the performance of binary classification models, particularly in distinguishing between positive and negative classes across various threshold settings. Typically, AUROC is defined as the area under the receiver operating characteristic curve, which plots the true positive rate (TPR) against the false positive rate (FPR) as the decision threshold is varied. Mathematically, AUROC is expressed as
\begin{align}
    \text{AUROC} &= \int_0^1 \text{TPR}(t) d\text{FPR}(t), \\
    \text{TPR} &= \frac{TP}{TP+FN}, \\ 
    \text{FPR} &= \frac{FP}{FP+TN}.
\end{align}
%where $TPR$ and $FPR$ are true positive rate and false positive rate.

For binary and multi-class lesion segmentation tasks, we utilized the dice similarity coefficient (DSC) for evaluation. DSC is useful for assessing the accuracy of a model in scenarios where spatial overlap between predicted and true masks is of primary interest. It is defined as
\begin{equation}
    \text{DSC} = \frac{2 \cdot |P \cap T|}{|P|+|T|},
\end{equation}
where $P$ represents the predicted segmentation and $T$ represents the ground truth.

In addition to DSC, we used the Hausdorff distance (HD) to evaluate the binary segmentation models. HD is a critical metric for image segmentation tasks, which measures the maximum discrepancy between the boundaries of the predicted segmentation and the ground truth. It evaluates the worst-case scenario by identifying the greatest distance from any point on one boundary to the closest point on the other boundary. Mathematically, the HD between two prediction segmentation $P$ and ground truth $T$ is defined as
\begin{equation}
    \text{HD} = \mathtt{max}\left\{\mathop{\mathtt{max}}\limits_{p\in P}\{\mathop{\mathtt{min}}\limits_{t\in T}\}||p-t||, \mathop{\mathtt{max}}\limits_{t\in T}\{\mathop{\mathtt{min}}\limits_{p\in P}\}||t-p|| \right\}.
\end{equation}

For pubic symphysis-fetal head tasks, the measurement of the angle of progression was conducted by constructing two lines from three specific landmarks. Firstly, we identified the two furthest points on the pubic symphysis contour based on the segmented image. Then, we drew a tangent line through the rightmost point of the pubic symphysis to define the fetal head region. The tangent line on the right side of the image, intersecting with the fetal head region, determines the third point for calculating the angle of progression (AoP). Finally, the angle formed by these three points constituted AoP.
The performance of AoP prediction was evaluated using the mean absolute error (MAE). MAE provides a straightforward interpretation of how far, on average, the predicted values deviate from the actual observed values, which is given by
\begin{equation}
    \text{MAE} = \frac{1}{n}\sum_{i=1}^n |y_i - \hat{y}_i|,
\end{equation}
where $n$ is the number of samples, $y_i$ is the actual value for the $i$-th sample, $\hat{y}_i$ is the predicted value for $i$-th sample. 

{
To evaluate the prediction fairness of the method, we adopted equal accuracy (EA) as the fairness evaluation metric, which measures the maximum gap in prediction accuracy between different groups (e.g., different hospitals, age groups):
\begin{equation}
    EA=\max_{k}\left|\operatorname{Score}\left(\mathcal{A}_{k}\right)-\overline{\operatorname{Score}}\right|.
\end{equation}
Here, $A_k$ represents the k-th client, $\operatorname{Score}(A_k)$ denotes the AUROC score of the test set for client k, and $ \overline{\operatorname{Score}}$ is the average score across all clients. Minimizing EA (i.e., narrowing the performance gap between groups) indicates achieving maximum fairness. 
}

The models for all tasks were trained using five different random seeds to determine the shuffling of the training data. We calculated the mean and standard deviation of the performance over the three iterations and computed the standard error, i.e., the standard deviation divided by the square root of 5. We obtained the 95\% confidence interval (CI) by multiplying the standard error by 1.96. Moreover, to determine whether there were significant differences, we performed two-sided t-tests between the significance of UltraFedFM compared to other methods.

% implementation details
\subsection*{Implementation details}
%-----------------------------------------------------------% 
{
In the pre-training stage, the image encoder of UltraFedFM is implemented by a basic vision Transformer33 (ViT-base) with 12 Transformer blocks and an embedding vector size of 768, whereas the decoder is a small vision Transformer (ViT-small) with 8 Transformer blocks and an embedding vector size of 512. The masking ratio is configured to 0.75, with an input size of $224 \times 224$. The model was pre-trained for 600 communication rounds (epochs) with a batch size of 512, and the warm-up period is 60 epochs. The local model was trained with 1 epoch in each communication round. We employed the AdamW optimizer with $\beta_1=0.9, \beta_2=0.95$, an initial learning rate of $1.5e-4$, and a weight decay of 0.05. 
For the fine-tuning stage, the input images are resized to $224 \times 224$, and Random rotations, flips, and crops were used as data augmentation. The batch size was set as 16. All models were trained with 100 epochs using the AdamW optimizer with a cosine learning rate scheduler. The first 10 epochs were used for learning rate warm-up. The drop-path probability was set to 0.1. The detailed configurations are listed in the Supplementary Table 4 and Supplementary Table 5.
}
\subsection*{Computing hardware and software}
We used Python (version 3.7.4) for all experiments and analyses in the study, which can be replicated using open-source libraries as outlined below. 
For pre-training, we used 8 32-GB NVIDIA GeForce Tesla V100 GPUs configured for multi-GPU training using DistributedDataParallel (DDP) as implemented by the framework PyTorch (version 1.11.0, CUDA 11.3). 
For fine-tuning, we used 1 32-GB NVIDIA GeForce Tesla V100 GPU.
Pillow library (version 9.5.0) and opencv-python (version 4.7.0) libraries were used to read images, which were then converted to the base64 string format using Python.
Timm library (version 0.9.2), torchvision (version 0.12.0) and opencv-python were applied for image processing and loading during training. 
Einops library (version 0.6.1) was applied for tensor operations in modeling. 
For model evaluation, we use the torchmetrics library (version 1.3.2) and pycm library (4.0) for classification task evaluations, and the segmentation-models-pytorch library (version 0.3.3) for segmentation task evaluations. 
Numpy (version 1.23.2) and Pandas (version 2.2.2), were used in data collection, preprocessing and data analysis.

\subsection*{Data availability}

The publicly available dataset for pre-training can be accessed from: BUV (\href{https://github.com/jhl-Det/CVA-Net/tree/main}{https://github.com/jhl-Det/CVA-Net/tree/main}), CLUST (\href{https://clust.ethz.ch/data.html}{https://clust.ethz.ch/data.html}), EchoNet-Dynamic (\href{https://echonet.github.io/dynamic/}{https://echonet.github.io/dynamic/}), FETAL-PLANES (\href{https://zenodo.org/records/3904280}{https://zeno\\do.org/records/3904280}), TDSC-ABUS (\href{https://tdsc-abus2023.grand-challenge.org/Dataset/}{https://tdsc-abus2023.grand-challenge.org/Dataset/}), Leg-3D-US (\href{https://www.cs.cit.tum.de/camp/publications/leg-3d-us-dataset/}{https://www.cs.cit.tu\\m.de/camp/publications/leg-3d-us-dataset/}), Thyroid Ultrasound Cine-clip (\href{https://stanfordaimi.azurewebsites.net/datasets/a72f2b02-7b53-4c5d-963c-d7253220bfd5}{https://stanfordaimi.azurewebsites.net/datasets/a72f2\\b02-7b53-4c5d-963c-d7253220bfd5}), SYSU-FLL-CEUS (\href{https://github.com/lemondan/Focal-liver-lesions-dataset-in-CEUS}{https://github.com/lemondan/Focal-liver-lesions-dataset-in-CEUS}), CAMUS (\href{https://www.creatis.insa-lyon.fr/Challenge/camus/index.html}{https://www.creatis.insa-lyon.fr/Challenge/camus/index.html}), COVID-BLUES (\href{https://github.com/NinaWie/COVID-BLUES}{https://github.com/NinaWie/COVID-BLUES}), NerveUS (\href{https://www.kaggle.com/competitions/ultrasound-nerve-segmentation}{https://www.kaggle.com/competitions/ultrasound-nerve-segmentation}), LEPset (\href{https://zenodo.org/records/8041285}{https://zenodo.org/records/\\8041285}), FPUS (\href{https://github.com/bharathprabakaran/FPUS23?tab=readme-ov-file}{https://github.com/bharathprabakaran/FPUS23?tab=readme-ov-file}), GBUSV (\href{https://github.com/sbasu276/FocusMAE}{https://github.com/sbasu276/Fo\\cusMAE}). The publicly available datasets for downstream tasks can be accessed from: LEPset-labeled (\href{https://zenodo.org/records/8041285}{https://zenodo.org/records\\/8041285}), SYSU-FLL-CEUS-labeled (\href{https://github.com/lemondan/Focal-liver-lesions-dataset-in-CEUS}{https://github.com/lemondan/Focal-liver-lesions-dataset-in-CEUS}), GBCU (\href{https://gbc-iitd.github.io/data/gbcu}{https://gbc-iitd.github.io/data/gbcu}), BUSI (\href{https://www.kaggle.com/datasets/aryashah2k/breast-ultrasound-images-dataset}{https://www.kaggle.com/datasets/aryashah2k/breast-ultrasound-images-dataset}), BUV-labeled (\href{https://github.com/jhl-Det/CVA-Net/tree/main}{https://github.com/jhl-Det/CVA-Net/tree/main}), BUS-BRA (\href{https://zenodo.org/records/8231412}{https://zenodo.org/records/8231412}), BUS-UCLM (\href{https://data.mendeley.com/datasets/7fvgj4jsp7/3}{https://data.me\\ndeley.com/datasets/7fvgj4jsp7/3}), POCUS (\href{https://github.com/jannisborn/covid19_ultrasound}{https://github.com/jannisborn/covid19 ultrasound}), FETAL-PLANES-labeled (\href{https://zenodo.org/records/3904280}{https://zenodo.org/records/3904280}), HFUS (\href{https://data.mendeley.com/datasets/td8r3ty79b/1}{https://data.mendeley.com/datasets/td8r3ty79b/1}), NerveUS-labeled (\href{https://www.kaggle.com/competitions/ultrasound-nerve-segmentation}{https://www.\\kaggle.com/competitions/ultrasound-nerve-segmentation}), DDTI (\href{https://www.kaggle.com/datasets/dasmehdixtr/ddti-thyroid-ultrasound-images}{https://www.kaggle.com/datasets/dasmehdixtr/ddti-thyroid-ultrasound-images}), Thyroid Ultrasound Cine-clip labeled (\href{https://stanfordaimi.azurewebsites.net/datasets/a72f2b02-7b53-4c5d-963c-d7253220bfd5}{https://stanfordaimi.azurewebsites.net/datasets/a72f2b02-7b53-4c5d-963c-d7253220bfd5}), TG3k (\href{https://github.com/haifangong/TRFE-Net-for-thyroid-nodule-segmentation}{https://github.com/haifangong/TRFE-Net-for-thyroid-nodule-segmentation}), TN3k (\href{https://github.com/haifangong/TRFE-Net-for-thyroid-nodule-segmentation}{https://github.c\\om/haifangong/TRFE-Net-for-thyroid-nodule-segmentation}), LUMINOUS (\href{https://users.encs.concordia.ca/~impact/luminous-database/}{https://users.encs.concordia.ca/~impact/luminous-database/}), CardiacUDA (\href{https://www.kaggle.com/datasets/xiaoweixumedicalai/cardiacudc-dataset}{https://www.kaggle.com/datasets/xiaoweixumedicalai/cardiacudc-dataset}), JNU-IFM (\href{https://figshare.com/articles/dataset/JNU-IFM/14371652}{https://figshare.c\\om/articles/dataset/JNU-IFM/14371652}). The UltraFedFM private dataset consists of routinely collected healthcare data. Owing to its sensitive nature and the risk of reidentification, the dataset is subject to controlled access by means of a structured application process. Data access enquiries may be made by \href{https://forms.gle/sdS5uX5FJjFRcr74A}{Google form}. We will review and aim to respond in a few weeks. The pre-trained and fine-tined models, as well as source code for pre-training, fine-tuning, inference, and data preprocessing, can be accessed at \href{https://github.com/yuncheng97/UltraFedFM}{https://github.com/yuncheng97/UltraFedFM}

\subsection*{Acknowledgement}
% This work was supported by Shenzhen-Hong Kong Joint Funding No. SGDX20211123112401002 (awarded to Z.L.), NSFC with Grant No. 62293482 (awarded to S.C.),  the Basic Research Project No. HZQB-KCZYZ-2021067 of Hetao Shenzhen HK S\&T Cooperation Zone (awarded to S.C.), Shenzhen General Program No. JCYJ20220530143600001 (awarded to Z.L.), the Shenzhen Outstanding Talents Training Fund 202002 (awarded to S.C.), the Guangdong Research Project No.2017ZT07X152  and No. 2019CX01X104 (awarded to S.C.), the Guangdong Provincial Key Laboratory of Future Networks of Intelligence (Grant No. 2022B1212010001) (awarded to J.R., Z.L., S.C.), the Guangdong Provincial Key Laboratory of BigData Computing CHUK-Shenzhen (awarded to Z.L.), the NSFC 61931024\&12326610 (awarded to Z.L.), the Key Area R\&D Program of Guangdong Province with grant No. 2018B030338001 (awarded to Z.L., S.C.), the Shenzhen Key Laboratory of Big Data and Artificial intelligence (Grant No. ZDSYS201707251409055) (awarded to Z.L., S.C.),  by China Association for Science and Technology Youth Care Program (awarded to Z.L.), and by Tencent \& Huawei Open Fund (awarded to Z.L.).

This work was supported by NSFC with Grant No. 62293482 (awarded to S.C), by the Basic Research Project No. HZQB-KCZYZ-2021067 of Hetao Shenzhen-HK S\&T Cooperation Zone (awarded to S.C.), by NSFC with Grant No. 62573371 (awarded to Z.L.), by the Shenzhen-Hong Kong Joint Funding No. SGDX20211123112401002 (awarded to Z.L.), by the Shenzhen General Program No. JCYJ20220530143600001 (awarded to Z.L.), by the Shenzhen Outstanding Talents Training Fund 202002 (awarded to S.C.), by the Guangdong Research Project No.2017ZT07X152 and No. 2019CX01X104 (awarded to S.C.), by the Guangdong Provincial Key Laboratory of Future Networks of Intelligence (Grant No. 2022B1212010001) (awarded to J.R., S.C.), by the Guangdong Provincial Key Laboratory of BigData Computing CHUK-Shenzhen (awarded to Z.L.), by the NSFC 61931024\&12326610 (awarded to Z.L.), by the Key Area R\&D Program of Guangdong Province with grant No. 2018B030338001 (awarded to Z.L., S.C.), by the Shenzhen Key Laboratory of Big Data and Artificial Intelligence (Grant No. SYSPG20241211173853027) (awarded to Z.L., S.C.), by China Association for Science and Technology Youth Care Program (awarded to Z.L.), by the NSFC 62225113 (awarded to B.D.), by National Key Research and Development Program of China under Grants 2023YFC2705702 (awarded to B.D.) and by Tencent \& Huawei Open Fund (awarded to Z.L.).

\bibliography{reference}

\newpage

\captionsetup[figure]{name=Supplementary Figure}
\captionsetup[table]{name=Supplementary Table}
% 在需要重新编号的位置插入以下代码
\setcounter{figure}{0}  % 重置计数器为0（显示为1）
\setcounter{table}{0}  % 重置计数器为0（显示为1）

\begin{table}[!t]
    \small
    \centering
    \caption{Summary of the pre-training datasets and the data distributions of all local clients.}
    \label{tab:pre-train}
    % \resizebox{\columnwidth}{!}{
    \renewcommand\tabcolsep{6pt}{
    \begin{tabular}{|c|c|c|c|c|c|}
    \hline
    \textbf{Clients} & \textbf{Name} & \textbf{Location} & \textbf{Main US Modality} & \textbf{Main Organ} & \textbf{Numbers} \\ \hline
    Client 1 & BUV \cite{buv} & Hong Kong, China & Superficial & Breast & 25,727 \\ \hline
    Client 2 & CLUST \cite{liverus-1, liverus-2}& Switzerland & Abdominal & Liver & 35,626 \\ \hline
    Client 3 & EchoNet-Dynamic \cite{echonet-dynamic}& California, USA & Echocardiogram & Heart & 150,243 \\ \hline
    Client 4 & FETAL-PLANES\cite{fetal_plane_1, fetal_plane_2}& Spain & Fetal & Fetal organs & 12,400 \\ \hline
    Client 5 & TDSC-ABUS \cite{3dbreastus} & Harbin, China & 3D Abdominal & Breast & 126,918 \\ \hline
    Client 6 & Leg-3D-US \cite{3dlegus} & German & 3D Muscle & Muscle & 115,263 \\ \hline
    Client 7 & Thyroid Ultrasound Cine-clip \cite{thyroidcineus}& California, USA & Superficial & Thyroid & 17,412 \\ \hline
    Client 8 & SYSU-FLL-CEUS \cite{ceus-1, ceus-2}& Guangzhou, China & Abdominal & Liver & 118,485 \\ \hline
    Client 9 & CAMUS \cite{camus}& France & Echocardiogram & Heart & 21,214 \\ \hline
    Client 10 & COVID-BLUES \cite{lungus} & United Kingdom & Lung & Lung & 42,226 \\ \hline
    Client 11 & NerveUS \cite{nus} & Georgia, USA & Superficial & Nerve & 11,143 \\ \hline
    Client 12 & LEPset \cite{LEPset}& Chongqing, China & Abdominal & Pancreas & 11,499 \\ \hline
    Client 13 & FPUS \cite{fpus} & Austria & Maternal-fetal & Fetal organs & 20,378 \\ \hline
    Client 14 & GBUSV \cite{gbusv} & India & Abdominal & Gallbladder & 10,553 \\ \hline
    Client 15 & ERUS & Sichuan, China & Endorectal & Colorectum & 63,426 \\ \hline
    Client 16 & PrivateUS & Sichuan, China & \begin{tabular}[c]{@{}c@{}}Abdominal, \\ Maternal-fetal, \\ Superficial,   \\ Echocardiogram\end{tabular} & \begin{tabular}[c]{@{}c@{}}Uterus, Vascular, Stomach,   \\ Spleen, Prostate, Testis, \\ Paratod Gland, Lymph Nodes\end{tabular} & 235,633 \\ \hline
    \end{tabular}}
    \label{tab:pretrain}
\end{table}

\begin{table}[!t]
    \centering
    \caption{Summary of the downstream tasks and the fine-tuning dataset distributions.}
    \label{tab:finetune}
    \resizebox{\columnwidth}{!}{
    \begin{tabular}{|cccccccc|}
\hline
\multicolumn{1}{|c|}{\textbf{Challenge}} & \multicolumn{1}{c|}{\textbf{Location}} & \multicolumn{1}{c|}{\textbf{Modality}} & \multicolumn{1}{c|}{\textbf{Organ}} & \multicolumn{1}{c|}{\textbf{Name}} & \multicolumn{1}{c|}{\textbf{Classes}} & \multicolumn{1}{c|}{\textbf{Train}} & \textbf{Test} \\ \hline
\multicolumn{8}{|c|}{\textbf{Classification}} \\ \hline
\multicolumn{1}{|c|}{\multirow{4}{*}{\begin{tabular}[c]{@{}c@{}}Malignant Cancer\\       Diagnose\end{tabular}}} & \multicolumn{1}{c|}{Chongqin, China} & \multicolumn{1}{c|}{Endoscopic} & \multicolumn{1}{c|}{Pancreas} & \multicolumn{1}{c|}{LEPset \cite{LEPset}} & \multicolumn{1}{c|}{\begin{tabular}[c]{@{}c@{}}Benign tumor,\\      Malignant tumor\end{tabular}} & \multicolumn{1}{c|}{2,800} & 700 \\ \cline{2-8} 
\multicolumn{1}{|c|}{} & \multicolumn{1}{c|}{Guangzhou, China} & \multicolumn{1}{c|}{\multirow{2}{*}{Abdominal}} & \multicolumn{1}{c|}{Liver} & \multicolumn{1}{c|}{SYSU-CEUS \cite{ceus-1, ceus-2}} & \multicolumn{1}{c|}{\begin{tabular}[c]{@{}c@{}}Hepatocellular carcinoma,\\      Hemangioma,\\      Focal nodular hyperplasia\end{tabular}} & \multicolumn{1}{c|}{10,910} & 2,728 \\ \cline{2-2} \cline{4-8} 
\multicolumn{1}{|c|}{} & \multicolumn{1}{c|}{India} & \multicolumn{1}{c|}{} & \multicolumn{1}{c|}{Gallbladder} & \multicolumn{1}{c|}{GBCU \cite{gbcu}} & \multicolumn{1}{c|}{\begin{tabular}[c]{@{}c@{}}Normal tumor,\\      Benign tumor,\\      Malignant tumor\end{tabular}} & \multicolumn{1}{c|}{2,292} & 573 \\ \cline{2-8} 
\multicolumn{1}{|c|}{} & \multicolumn{1}{c|}{Egypt} & \multicolumn{1}{c|}{\multirow{3}{*}{Superficial}} & \multicolumn{1}{c|}{\multirow{3}{*}{Breast}} & \multicolumn{1}{c|}{BUS \cite{busi}, BUV \cite{buv}} & \multicolumn{1}{c|}{\begin{tabular}[c]{@{}c@{}}Normal tumor,\\      Benign tumor,\\      Malignant tumor\end{tabular}} & \multicolumn{1}{c|}{1,119} & 300 \\ \cline{2-2} \cline{5-8} 
\multicolumn{1}{|c|}{} & \multicolumn{1}{c|}{Brazil} & \multicolumn{1}{c|}{} & \multicolumn{1}{c|}{} & \multicolumn{1}{c|}{BUS-BRA \cite{gomez2024bus-bra}} & \multicolumn{1}{c|}{\begin{tabular}[c]{@{}c@{}}Normal tumor,\\      Benign tumor,\\      Malignant tumor\end{tabular}} & \multicolumn{1}{c|}{851} & 213 \\ \cline{2-2} \cline{5-8} 
\multicolumn{1}{|c|}{} & \multicolumn{1}{c|}{Spain} & \multicolumn{1}{c|}{} & \multicolumn{1}{c|}{} & \multicolumn{1}{c|}{BUS-UCLM \cite{vallez2025bus-uclm}} & \multicolumn{1}{c|}{\begin{tabular}[c]{@{}c@{}}Normal tumor,\\      Benign tumor,\\      Malignant tumor\end{tabular}} & \multicolumn{1}{c|}{546} & 137 \\ \hline
\multicolumn{1}{|c|}{\begin{tabular}[c]{@{}c@{}}Cancer   Infiltration \\      depth staging\end{tabular}} & \multicolumn{1}{c|}{Sichuan, China} & \multicolumn{1}{c|}{Endorectal} & \multicolumn{1}{c|}{Colorectum} & \multicolumn{1}{c|}{ERUS} & \multicolumn{1}{c|}{\begin{tabular}[c]{@{}c@{}}Colorectal tumor staging: \\ T1,T2,T3,T4,T5\end{tabular}} & \multicolumn{1}{c|}{1,239} & 304 \\ \hline
\multicolumn{1}{|c|}{\multirow{2}{*}{\begin{tabular}[c]{@{}c@{}}Organ disease \\      diagnosis\end{tabular}}} & \multicolumn{1}{c|}{Sichuan, China} & \multicolumn{1}{c|}{Echocardiogram} & \multicolumn{1}{c|}{Heart} & \multicolumn{1}{c|}{EUS} & \multicolumn{1}{c|}{\begin{tabular}[c]{@{}c@{}}Patent ductus arteriosus, \\      Atrial septal defect, \\      Ventricular septal defect,\\      Heart valve disease\end{tabular}} & \multicolumn{1}{c|}{10,368} & 2,592 \\ \cline{2-8} 
\multicolumn{1}{|c|}{} & \multicolumn{1}{c|}{Sichuan, China} & \multicolumn{1}{c|}{Abdominal} & \multicolumn{1}{c|}{Kidney} & \multicolumn{1}{c|}{KUS} & \multicolumn{1}{c|}{\begin{tabular}[c]{@{}c@{}}Normal,\\      Cyst,\\      Hydronephrosis,\\      Dysplasia,\\      Benign tumor,\\      Malignant tumor,\end{tabular}} & \multicolumn{1}{c|}{5,662} & 1,415 \\ \hline
\multicolumn{1}{|c|}{Covid-19 prediction} & \multicolumn{1}{c|}{United Kingdom} & \multicolumn{1}{c|}{Lung} & \multicolumn{1}{c|}{Lung} & \multicolumn{1}{c|}{POCUS \cite{pocus}} & \multicolumn{1}{c|}{\begin{tabular}[c]{@{}c@{}}Normal,\\      COVID-19, \\      Bacterial pneumonia,\\      Viral pneumonia\end{tabular}} & \multicolumn{1}{c|}{3,755} & 939 \\ \hline
\multicolumn{1}{|c|}{\begin{tabular}[c]{@{}c@{}}Fetal   plane \\      recognition\end{tabular}} & \multicolumn{1}{c|}{Spain} & \multicolumn{1}{c|}{Maternal-fetal} & \multicolumn{1}{c|}{Fetal Organ} & \multicolumn{1}{c|}{FETAL PLANES \cite{fetal_plane_1, fetal_plane_2}} & \multicolumn{1}{c|}{\begin{tabular}[c]{@{}c@{}}Abdomen,\\      Brain,\\      Femur,\\       Thorax,\\      Maternal cervix,\\      Other\end{tabular}} & \multicolumn{1}{c|}{15,623} & 3,906 \\ \hline
\multicolumn{1}{|c|}{\begin{tabular}[c]{@{}c@{}}Ultrasound imaging \\      quality assessment\end{tabular}} & \multicolumn{1}{c|}{Poland} & \multicolumn{1}{c|}{High-frequency} & \multicolumn{1}{c|}{Skin} & \multicolumn{1}{c|}{HFUS \cite{hfus}} & \multicolumn{1}{c|}{\begin{tabular}[c]{@{}c@{}}Good skin quality,\\      Bad skin quality\end{tabular}} & \multicolumn{1}{c|}{13,940} & 3,485 \\ \hline
\multicolumn{8}{|c|}{\textbf{Segmentation}} \\ \hline
\multicolumn{1}{|c|}{\begin{tabular}[c]{@{}c@{}}Critical   organ \\      segmentation\end{tabular}} & \multicolumn{1}{c|}{Georgia,   USA} & \multicolumn{1}{c|}{\multirow{2}{*}{Superficial}} & \multicolumn{1}{c|}{Brachial plexus} & \multicolumn{1}{c|}{NUS \cite{nus}} & \multicolumn{1}{c|}{\begin{tabular}[c]{@{}c@{}}Brachial plexus,\\      Background\end{tabular}} & \multicolumn{1}{c|}{1,877} & 446 \\ \cline{1-2} \cline{4-8} 
\multicolumn{1}{|c|}{Lesion segmentation} & \multicolumn{1}{c|}{Guangzhou, China} & \multicolumn{1}{c|}{} & \multicolumn{1}{c|}{Thyroid} & \multicolumn{1}{c|}{\begin{tabular}[c]{@{}c@{}}DDTI \cite{ddti}, \\ Thyroid Cine-Clip \cite{thyroidcineus},\\ TG3k \cite{tg3k}, \\ TN3k \cite{tn3k-1, tn3k-2}\end{tabular}} & \multicolumn{1}{c|}{\begin{tabular}[c]{@{}c@{}}Thyroid nodule,\\      Background\end{tabular}} & \multicolumn{1}{c|}{20,109} & 5,027 \\ \hline
\multicolumn{1}{|c|}{Muscle segmentation} & \multicolumn{1}{c|}{Montreal, Canada} & \multicolumn{1}{c|}{Muscle} & \multicolumn{1}{c|}{\begin{tabular}[c]{@{}c@{}}Lumbar \\      multifidus \\      muscle\end{tabular}} & \multicolumn{1}{c|}{LUMINOUS \cite{luminous}} & \multicolumn{1}{c|}{\begin{tabular}[c]{@{}c@{}}Muscle,\\      Background\end{tabular}} & \multicolumn{1}{c|}{276} & 65 \\ \hline
\multicolumn{1}{|c|}{Heart ventricle segmentation} & \multicolumn{1}{c|}{Guangzhou, China} & \multicolumn{1}{c|}{Echocardiogram} & \multicolumn{1}{c|}{Heart} & \multicolumn{1}{c|}{CardiacUDA \cite{cardiacuda}} & \multicolumn{1}{c|}{\begin{tabular}[c]{@{}c@{}} Left ventricle \\  Right ventricle \\ pulmonary artery  \\ left atrium \\ right atrium \\Background \end{tabular}} & \multicolumn{1}{c|}{1904} & 353 \\ \hline
\multicolumn{1}{|c|}{\begin{tabular}[c]{@{}c@{}}Fetal   head descent\\      \&\\      Progression \\      assessment\end{tabular}} & \multicolumn{1}{c|}{Wuhan, China} & \multicolumn{1}{c|}{Transperineal} & \multicolumn{1}{c|}{\begin{tabular}[c]{@{}c@{}}Pubic\\      \& \\      Fetal head\end{tabular}} & \multicolumn{1}{c|}{JNU-IFM \cite{jnu}} & \multicolumn{1}{c|}{\begin{tabular}[c]{@{}c@{}}Pubic symphysis,\\      Fetal head,\\      Background\end{tabular}} & \multicolumn{1}{c|}{2,571} & 643 \\ \hline
\end{tabular}
    }
\end{table}

% \usepackage{graphicx}
\begin{table}[tbh]
\centering
\caption{Summary of the downstream organ-agnostic dataset distributions}
\label{tab:organ-agnostic}
\renewcommand\tabcolsep{5pt}
% \resizebox{\textwidth}{!}{%
% \setlength{\tabcolsep}{1.5mm}{
\begin{tabular}{|c|c|c|c|c|}
\hline
\textbf{Task} & \textbf{Classes} & \textbf{Name} & \textbf{Train} & \textbf{Test} \\ \hline
\begin{tabular}[c]{@{}c@{}}Organ-agnostic\\ disease diagnosis\end{tabular} & \begin{tabular}[c]{@{}c@{}}Pancreatic malignant tumor\\ Gallbladder malignant tumor\\ Liver hepatocelular carcinoma\\ Breast malignant tumor\\ Kidney malignant tumor\\ Colorectal T4-stage tumor\\ Cardiac valve disease\\ Lung Covid-19\end{tabular} & \begin{tabular}[c]{@{}c@{}} LEPset\cite{LEPset} \\ GBCU\cite{gbcu}\\ SYSU-CEUS\cite{ceus-1} \\ BUS\cite{busi}, BUV\cite{buv} \\ KUS \\ ERUS \\ EUS \\ POCUS \cite{pocus} \end{tabular} & 11,670 & 3,177 \\ \hline
\begin{tabular}[c]{@{}c@{}}Organ-agnostic\\ lesion segmentation\end{tabular} & \begin{tabular}[c]{@{}c@{}}Thyroid Nodule\\ Brachial plexus\\ Muscle\\ Heart ventricle \\ Background\end{tabular} & \begin{tabular}[c]{@{}c@{}} Thyroid Cine-clip \cite{thyroidcineus}, DDTI\cite{ddti}, TG3k\cite{tg3k}, TN3k\cite{tn3k-1}\\ NUS \cite{nus} \\LUMINOUS \cite{luminous} \\ CardiacUDA \cite{cardiacuda} \end{tabular}  & 44,506 & 11,076 \\ \hline
\end{tabular}%
% }
\end{table}

\begin{table}[h]
\centering
\caption{UltraFedFM pre-training settings. UltraFedFM uses ViT-B/16 as the encoder by default.}
\label{tab:pretraing_config}
% \renewcommand\tabcolsep{12mm}
% \resizebox{\textwidth}{!}{
\begin{tabular}{ll}
\toprule
  Config & Value  \\
\midrule
    masking ratio & 75\% \\
    input size & 224 $\times$ 224 \\
    optimizer & AdamW \\
    base learning rate & 1.5e-4 \\
    weight decay & 0.05 \\
    optimizer momentum & $\beta_1$, $\beta_2$ = 0.9, 0.95 \\
    batch size & 512 \\
    learning rate schedule & cosine decay \\
    layer-wise learning rate decay & 1.0 \\
    warmup epochs & 60 \\
    total pre-training epochs & 600 \\
    local pre-training epochs & 1 \\
    augmentation & RandomResizedCrop \\
\bottomrule
\end{tabular}
% }
\end{table}

\begin{table}[h]
\centering
\caption{UltraFedFM fine-tuning settings. UltraFedFM uses ViT-B/16 as the encoder by default.}
\label{tab:finetune_config}
\begin{tabular}{ll}
\toprule
  Config & Value  \\
\midrule
    input size & 224 $\times$ 224 \\
    optimizer & AdamW \\
    base learning rate & 5e-4 \\
    weight decay & 0.05 \\
    optimizer momentum & $\beta_1$, $\beta_2$ = 0.9, 0.95 \\
    batch size & 16 \\
    learning rate schedule & cosine decay \\
    layer-wise learning rate decay & 0.75 \\
    warmup epochs & 10 \\
    total fine-tuning epochs & 100 \\
    augmentation & Random rotation, Random flip, Random crop size \\
    drop path & 0.1 \\
\bottomrule
\end{tabular}
\end{table}

% \usepackage{graphicx}
\begin{table}[!t]
\centering
\caption{Specific quantitative results of different self-supervised methods}
\label{tab:ssl}
\renewcommand\tabcolsep{19pt}
% \resizebox{\textwidth}{!}{%
\begin{tabular}{ccccccc}
\toprule
 & \multicolumn{1}{c}{SimCLR} & \multicolumn{1}{c}{SwAV} & \multicolumn{1}{c}{MoCo} & \multicolumn{1}{c}{DINO} & \multicolumn{1}{c}{MAE} & UltraFedFM \\ \midrule
 & \multicolumn{6}{c}{Classification (AUROC)} \\ \midrule
Pancreas & \multicolumn{1}{c}{0.862} & \multicolumn{1}{c}{0.854} & \multicolumn{1}{c}{0.871} & \multicolumn{1}{c}{0.886} & \multicolumn{1}{c}{0.874} & 0.914 \\ 
Gallbladder & \multicolumn{1}{c}{0.867} & \multicolumn{1}{c}{0.852} & \multicolumn{1}{c}{0.870} & \multicolumn{1}{c}{0.879} & \multicolumn{1}{c}{0.881} & 0.957 \\ 
Liver & \multicolumn{1}{c}{0.664} & \multicolumn{1}{c}{0.622} & \multicolumn{1}{c}{0.693} & \multicolumn{1}{c}{0.704} & \multicolumn{1}{c}{0.719} & 0.764 \\ 
Breast & \multicolumn{1}{c}{0.858} & \multicolumn{1}{c}{0.851} & \multicolumn{1}{c}{0.872} & \multicolumn{1}{c}{0.885} & \multicolumn{1}{c}{0.881} & 0.922 \\ 
Colon & \multicolumn{1}{c}{0.827} & \multicolumn{1}{c}{0.811} & \multicolumn{1}{c}{0.836} & \multicolumn{1}{c}{0.845} & \multicolumn{1}{c}{0.852} & 0.900 \\ 
Lung & \multicolumn{1}{c}{0.938} & \multicolumn{1}{c}{0.912} & \multicolumn{1}{c}{0.944} & \multicolumn{1}{c}{0.954} & \multicolumn{1}{c}{0.961} & 0.987 \\ 
Heart & \multicolumn{1}{c}{0.887} & \multicolumn{1}{c}{0.865} & \multicolumn{1}{c}{0.91} & \multicolumn{1}{c}{0.915} & \multicolumn{1}{c}{0.922} & 0.973 \\ 
Fetal & \multicolumn{1}{c}{0.917} & \multicolumn{1}{c}{0.905} & \multicolumn{1}{c}{0.943} & \multicolumn{1}{c}{0.952} & \multicolumn{1}{c}{0.984} & 0.996 \\ \midrule
\multicolumn{1}{l}{} & \multicolumn{6}{c}{Segmentation (DSC)} \\ \midrule
Muscle & \multicolumn{1}{c}{0.817} & \multicolumn{1}{c}{0.804} & \multicolumn{1}{c}{0.853} & \multicolumn{1}{c}{0.860} & \multicolumn{1}{c}{0.874} & 0.902 \\ 
Thyroid & \multicolumn{1}{c}{0.781} & \multicolumn{1}{c}{0.773} & \multicolumn{1}{c}{0.792} & \multicolumn{1}{c}{0.800} & \multicolumn{1}{c}{0.812} & 0.879 \\ 
Nerve & \multicolumn{1}{c}{0.746} & \multicolumn{1}{c}{0.733} & \multicolumn{1}{c}{0.747} & \multicolumn{1}{c}{0.752} & \multicolumn{1}{c}{0.762} & 0.790 \\ 
Fetal head & \multicolumn{1}{c}{0.725} & \multicolumn{1}{c}{0.704} & \multicolumn{1}{c}{0.773} & \multicolumn{1}{c}{0.804} & \multicolumn{1}{c}{0.810} & 0.842 \\ 
Heart & \multicolumn{1}{c}{0.704} & \multicolumn{1}{c}{0.677} & \multicolumn{1}{c}{0.725} & \multicolumn{1}{c}{0.737} & \multicolumn{1}{c}{0.744} & 0.768 \\ \bottomrule
\end{tabular}%
% }
\end{table}

% \usepackage{graphicx}
\begin{table}[tbh]
\centering
\caption{Average accuracy score of UltraFedFM across 8 diseases, and comparison with the results of seven ultrasonograph clinicians}
\label{tab:human}
\resizebox{\textwidth}{!}{%
\begin{tabular}{lcccccccc}
\toprule
Disease class & UltraFedFM & Clinician-A & Clinician-B & Clinician-C & Clinician-D & Clinician-E & Clinician-F & Clinician-G \\ \midrule
Breast malignant tumor & 0.900 & 0.600 & 1.000 & 0.700 & 0.800 & 0.800 & 0.700 & 1.000 \\ 
Kidney malignant tumor & 1.000 & 0.900 & 0.800 & 1.000 & 0.800 & 0.800 & 1.000 & 0.800 \\ 
Liver hepatocellular carcinoma & 1.000 & 0.833 & 0.750 & 0.916 & 0.500 & 0.833 & 0.833 & 0.750 \\ 
Gallbladder malignant tumor & 0.800 & 0.100 & 0.200 & 0.300 & 0.600 & 0.400 & 0.100 & 0.500 \\ 
Pancreatic malignant tumor & 0.875 & 0.250 & 0.250 & 0.375 & 0.375 & 0.625 & 0.375 & 0.375 \\ 
Colorectal T4 stage tumor & 0.900 & 0.800 & 0.600 & 0.700 & 0.700 & 0.600 & 0.600 & 0.500 \\ 
Cardiac valve disease & 0.700 & 0.800 & 0.600 & 0.600 & 0.700 & 0.700 & 0.800 & 0.700 \\ 
Lung Covid-19 & 0.900 & 0.900 & 0.700 & 0.700 & 0.700 & 0.700 & 0.700 & 0.900 \\ \bottomrule
\end{tabular}%
}
\end{table}

\begin{table}[h]
    \centering
    \caption{Summary of Models and Publications for Fetal Plane and Gallbladder Cancer Classification}
    \label{tab:models_summary1}
    \begin{tabular}{|l|l|l|p{4cm}|l|}
        \hline
        Task & Dataset & Models & Published journal & Published time \\
        \hline
        \multirow{7}{*}{Fetal plane classification} 
         & \multirow{7}{*}{Fetal Plane \cite{fetal_plane_1, fetal_plane_2}} 
            & ResNet \cite{he2016resnet}& Conference on Computer Vision and Pattern Recognition & 2015 \\
         &  & SonoNet \cite{baumgartner2017sononet}& Transactions on Medical Imaging & 2017 \\
         &  & ViT \cite{vit}& International Conference on Learning Representations & 2020 \\
         &  & Zhang et al. \cite{zhang2021automatic}& Medicine & 2021 \\
         &  & HCN \cite{tanwar2024hcn}& International Journal of Imaging Systems and Technology & 2024 \\
         &  & Krishna et al. \cite{krishna2024standard}& International Journal of Imaging Systems and Technology & 2024 \\
         &  & UltraFedFM & - & - \\
        \hline
        \multirow{7}{*}{Gallbladder cancer classification} 
         & \multirow{7}{*}{GBCU \cite{gbcu}} 
            & ResNet \cite{he2016resnet}& Conference on Computer Vision and Pattern Recognition & 2015 \\
         &  & ViT \cite{vit}& International Conference on Learning Representations & 2020 \\
         &  & GBCNet \cite{gbcu}& Conference on Computer Vision and Pattern Recognition & 2022 \\
         &  & Radformer \cite{basu2023radformer}& Medical Image Analysis & 2023 \\
         &  & Focusmae \cite{gbusv}& Conference on Computer Vision and Pattern Recognition & 2024 \\
         &  & GBCHV \cite{hasan2025gbchv}& Scientific Reports & 2025 \\
         &  & UltraFedFM & - & - \\
         \hline
    \end{tabular}
\end{table}

\begin{table}[h]
    \centering
    \caption{Summary of Models and Publications for Fetal Plane and Gallbladder Cancer Classification}
    \label{tab:models_summary2}
    \begin{tabular}{|l|l|l|p{4cm}|l|}
        \hline
        Task & Dataset & Models & Published journal & Published time \\
        \hline
        \multirow{7}{*}{Breast nodule segmentation} 
         & \multirow{7}{*}{BUSI \cite{busi}} 
         & UNet \cite{unet}& International Conference on Medical Image Computing and Computer-Assisted Intervention & 2015 \\
         &  & TransUNet \cite{chen2021transunet}& ArXiv & 2021 \\
         &  & nnUNet \cite{isensee2021nnunet}& Nature Methods & 2021 \\
         &  & AAU-Net \cite{chen2022aaunet}& Transactions on Medical Imaging & 2023 \\
         &  & EH-former \cite{qu2024ehformer}& Information Fusion & 2024 \\
         &  & FABRF-Net \cite{liu2025fabrfnet}& Information Fusion & 2025 \\
         &  & EMGANet \cite{huang2025emganet}& Journal of Biomedical and Health Informatics & 2025 \\
         &  & UltraFedFM & - & - \\
         \hline
         \multirow{7}{*}{Thyroid nodule segmentation} 
         & \multirow{7}{*}{TN3K \cite{tn3k-1, tn3k-2}} 
            & UNet \cite{unet}& International Conference on Medical Image Computing and Computer-Assisted Intervention & 2015 \\
         &  & TransUNet \cite{chen2021transunet}& ArXiv & 2021  \\
         &  & nnUNet \cite{isensee2021nnunet}& Nature Methods & 2021 \\
         &  & MS-MTL \cite{basu2023radformer}& Medical Image Analysis & 2023 \\
         &  & TRFE \cite{gbusv}& Computers in biology and medicine & 2024 \\
         &  & SHAN \cite{hasan2025gbchv}& International Conference on Medical Image Computing and Computer-Assisted Intervention & 2025 \\
         &  & UltraFedFM & - & - \\
         \hline
    \end{tabular}
\end{table}

\begin{figure}[!t]
    \centering
    \includegraphics[width=\linewidth]{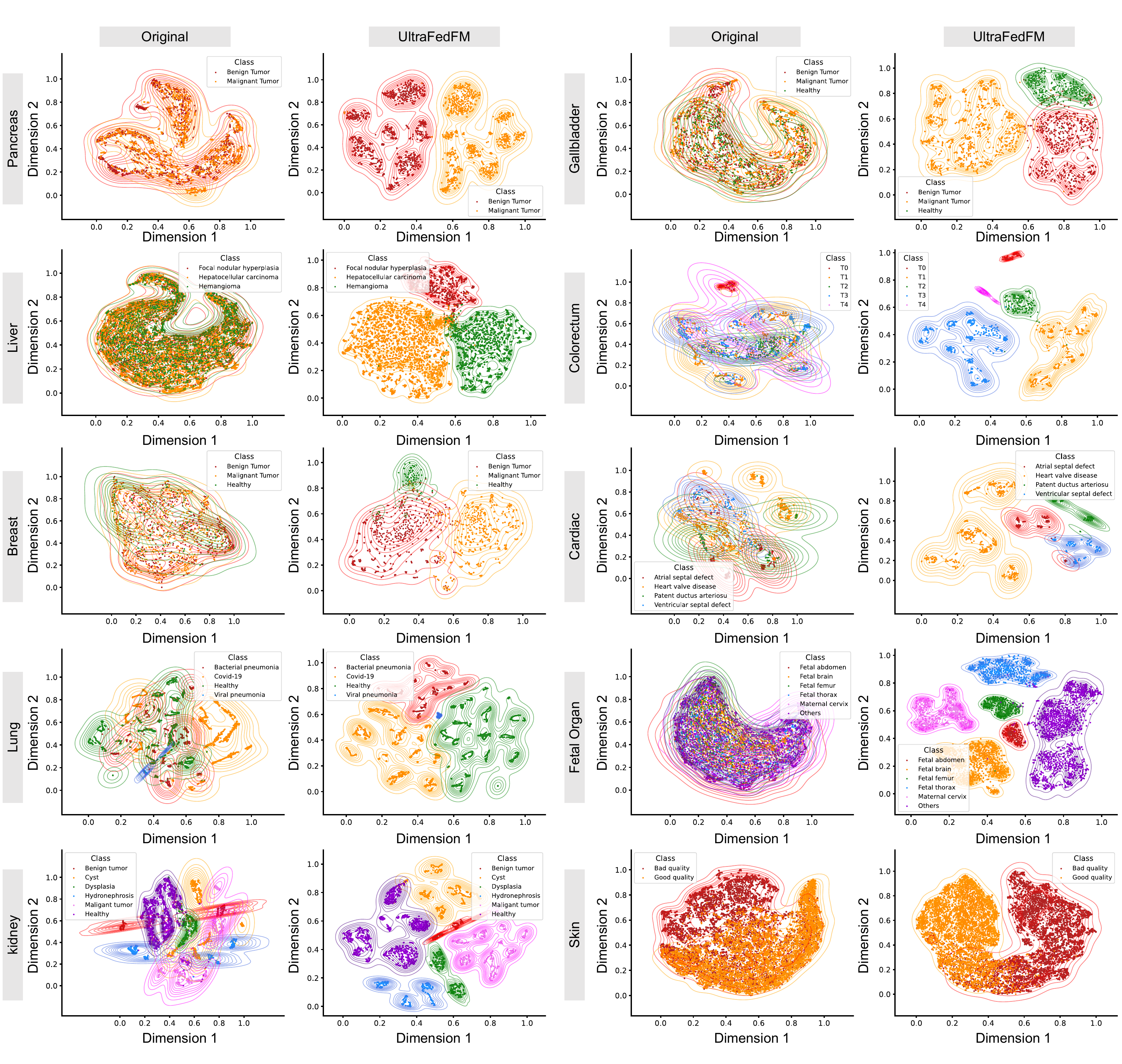}
    \caption{\textbf{Visualization of the features space.} Features from the foundation model and baseline model are extracted on the 10 independent disease diagnosis tasks using the tSNE dimensionality reduction approach. The x-axis corresponds to dimension 1, and the y-axis corresponds to dimension 2 of the reduced feature space. The density contours of each class are underlaid to highlight separability between classes in the feature space.}
    \label{supp:tSNE}
\end{figure}

\begin{figure}[!t]
    \centering
    \includegraphics[width=\linewidth]{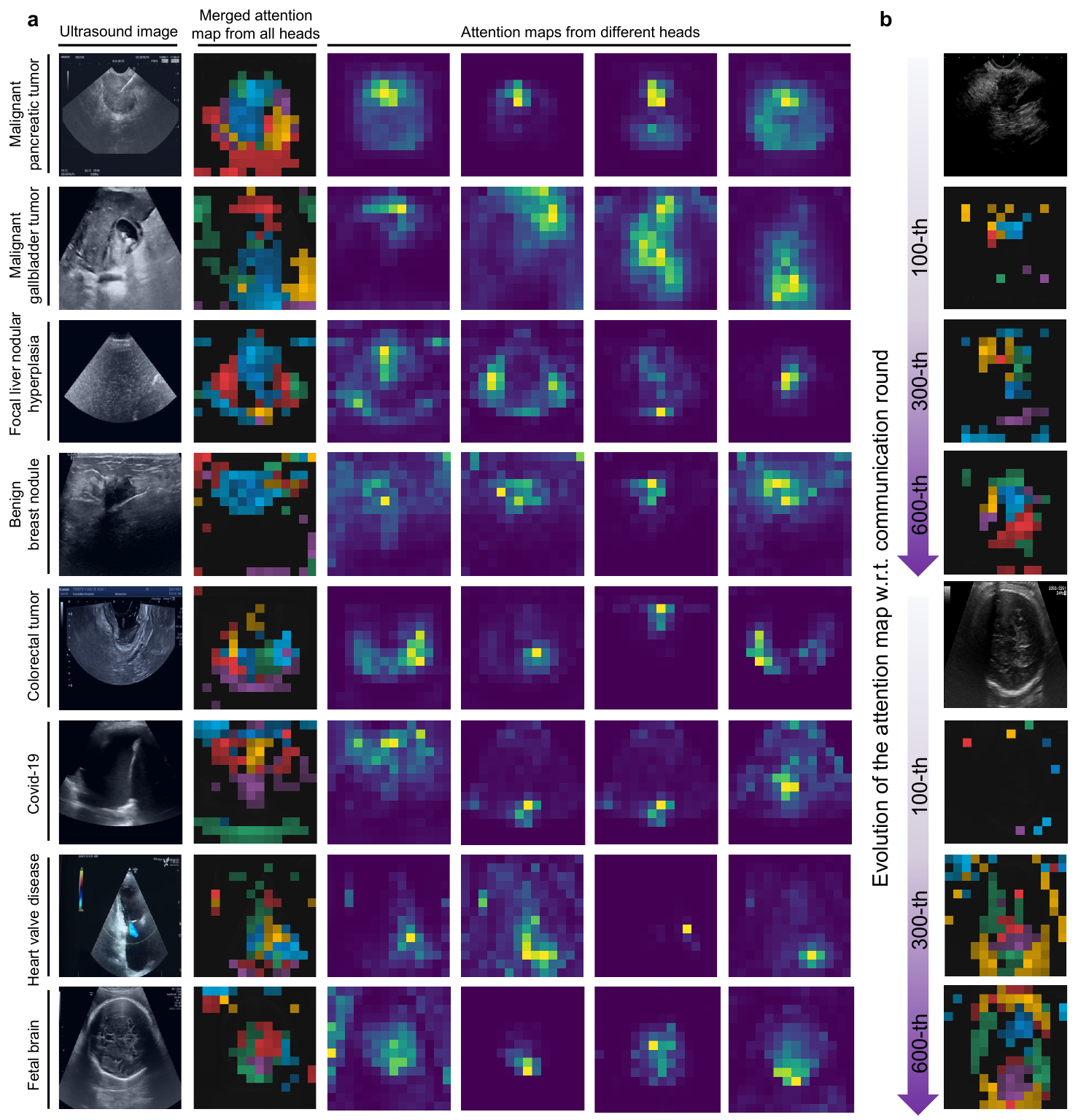}
    \caption{\textbf{Visualization of the attention map.} Explainability of the learned ultrasound image representations of UltraFedFM. \textbf{a} Attention maps of different heads of UltraFedFM on images of different organs are visualized. \textbf{b} Evolution of attention maps with respect to communication rounds during the federated pre-training.}
    \label{supp:attention}
\end{figure}

\begin{figure}[!t]
    \centering
    \includegraphics[width=\linewidth]{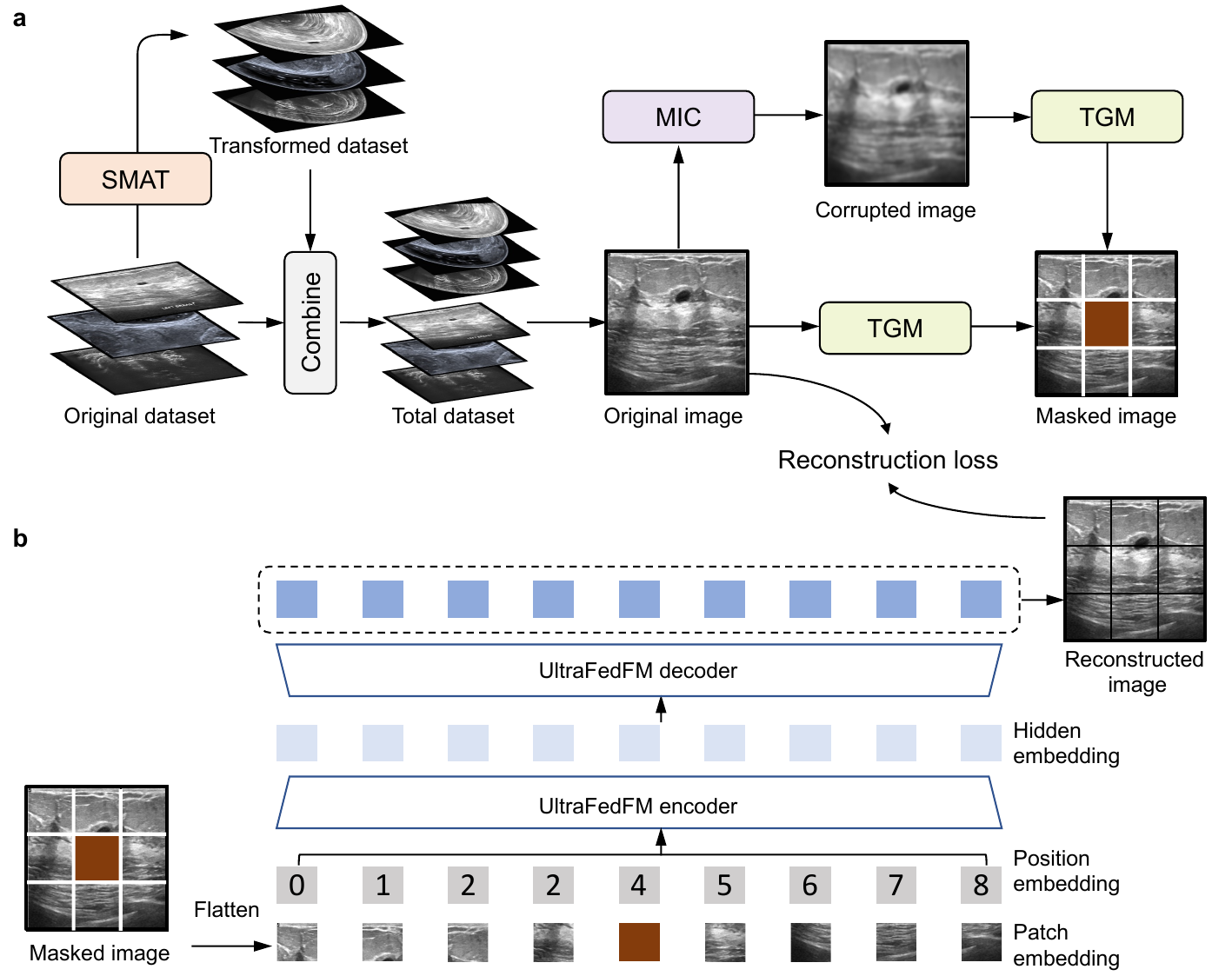}
    \caption{\textbf{UltraFedFM pre-training framework.} The local training at each client is based on a Masked Autoencoder (MAE) architecture, consisting of an encoder to extract image features and a decoder to perform pixel-level image reconstruction. \textbf{a} Ultrasound image masking (UIM). To address the unique challenge of ultrasound imaging, we proposed the ultrasound image masking incorporated three additional modules: scanning mode-aware transformation (SMAT), mixed image corruption (MIC), and texture-guided masking (TGM). Initially, the local data of each client undergoes SMAT to generate paired scanning mode images, which mimic various ultrasound modalities. Both real and simulated images are used during the training process. Throughout pre-training, each image may be applied to different quality corruptions through MIC, simulating real-world imaging artifacts. Subsequently, TGM generates a unique mask for each image. \textbf{b} Masked image modeling. The masked image is then passed through the encoder and decoder to produce a reconstructed image. Finally, the reconstruction loss is calculated by comparing the reconstructed image with the original input, ensuring that the model learns robust features from diverse and degraded ultrasound images.}
    \label{fig:framework}
\end{figure}

\begin{figure}[!t]
    \centering
    \includegraphics[width=\linewidth]{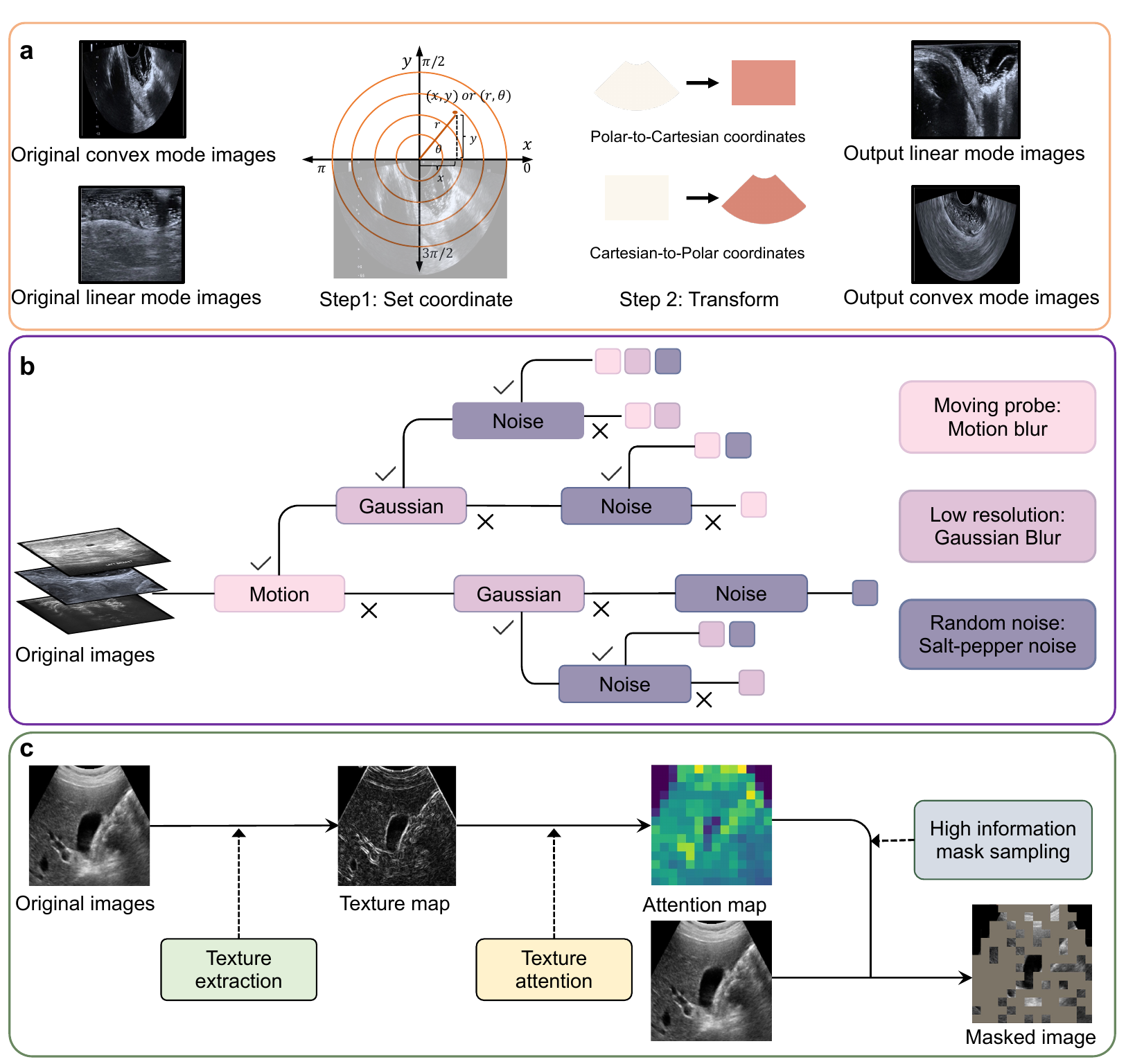}
    \caption{\textbf{Detailed illustration of the ultrasound masked image modeling (UltraMIM).} \textbf{a} Scanning mode-aware transformation (SMAT). To balance the proportion of different ultrasound scanning modes and generate extra training samples, we transform the original scanning mode into another mode based on the polar-cartesian corrodinates transformation. \textbf{b} Mixed image corruption (MIC). To enhance the robustness of the foundation model against multiple low-quality conditions in real practice, We add an additional de-corruption branch to the original reconstruction stage and use a mixture of three image corruption techniques to generate low-quality images. \textbf{c} Texture-guided masking (TGM). It is difficult to reconstruct areas with high information density in an image. We use the unique texture information of ultrasound images to mine high information density areas in the image, allowing the model to learn more meaningful information through reconstruction tasks.}
    \label{fig:AMIM}
\end{figure}

\begin{figure}[!t]
    \centering
    \includegraphics[width=\linewidth]{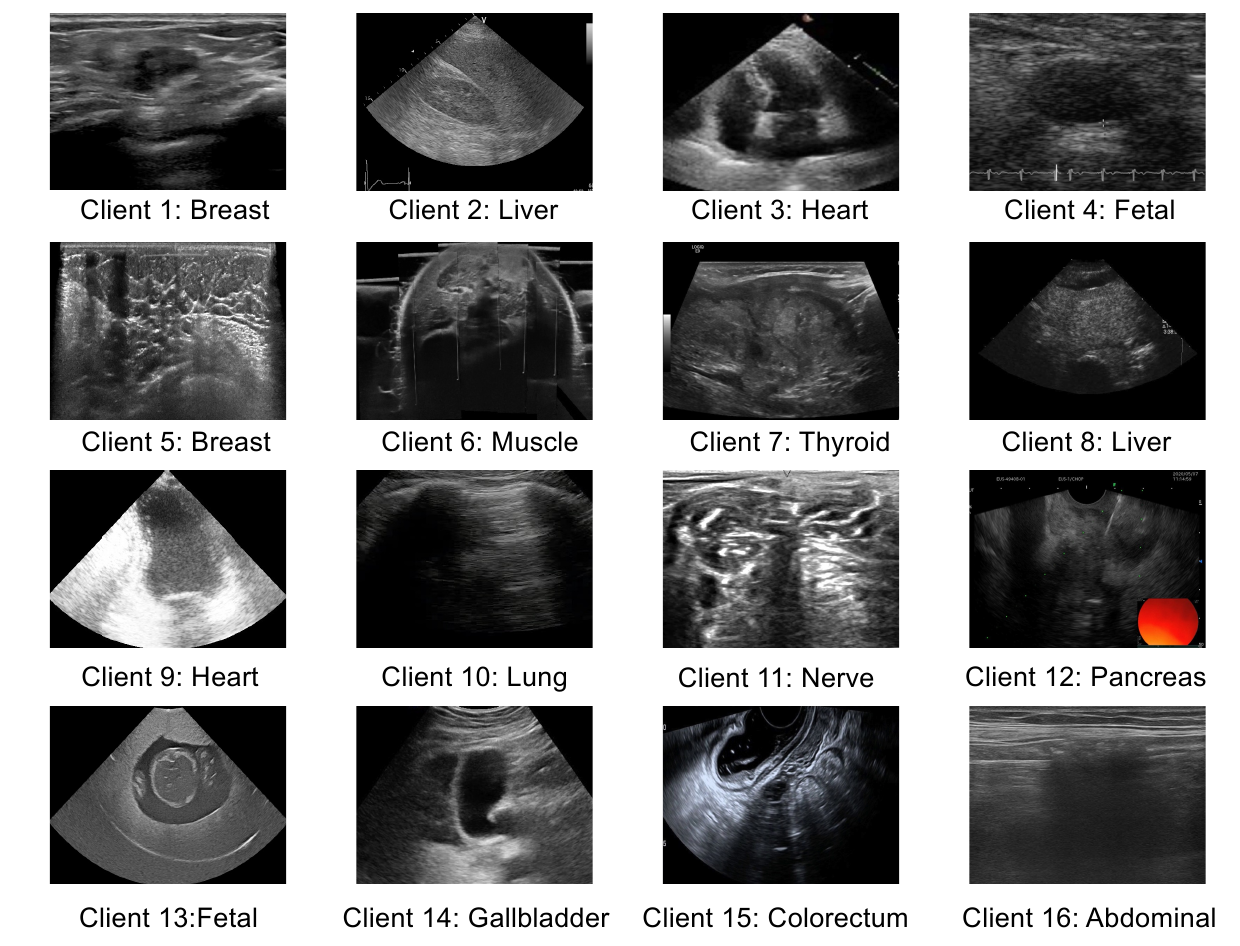}
    \caption{Sample images of each client in the pre-training dataset.}
    \label{fig:example_pretrain}
\end{figure}

\begin{figure}[!t]
    \centering
    \includegraphics[width=\linewidth]{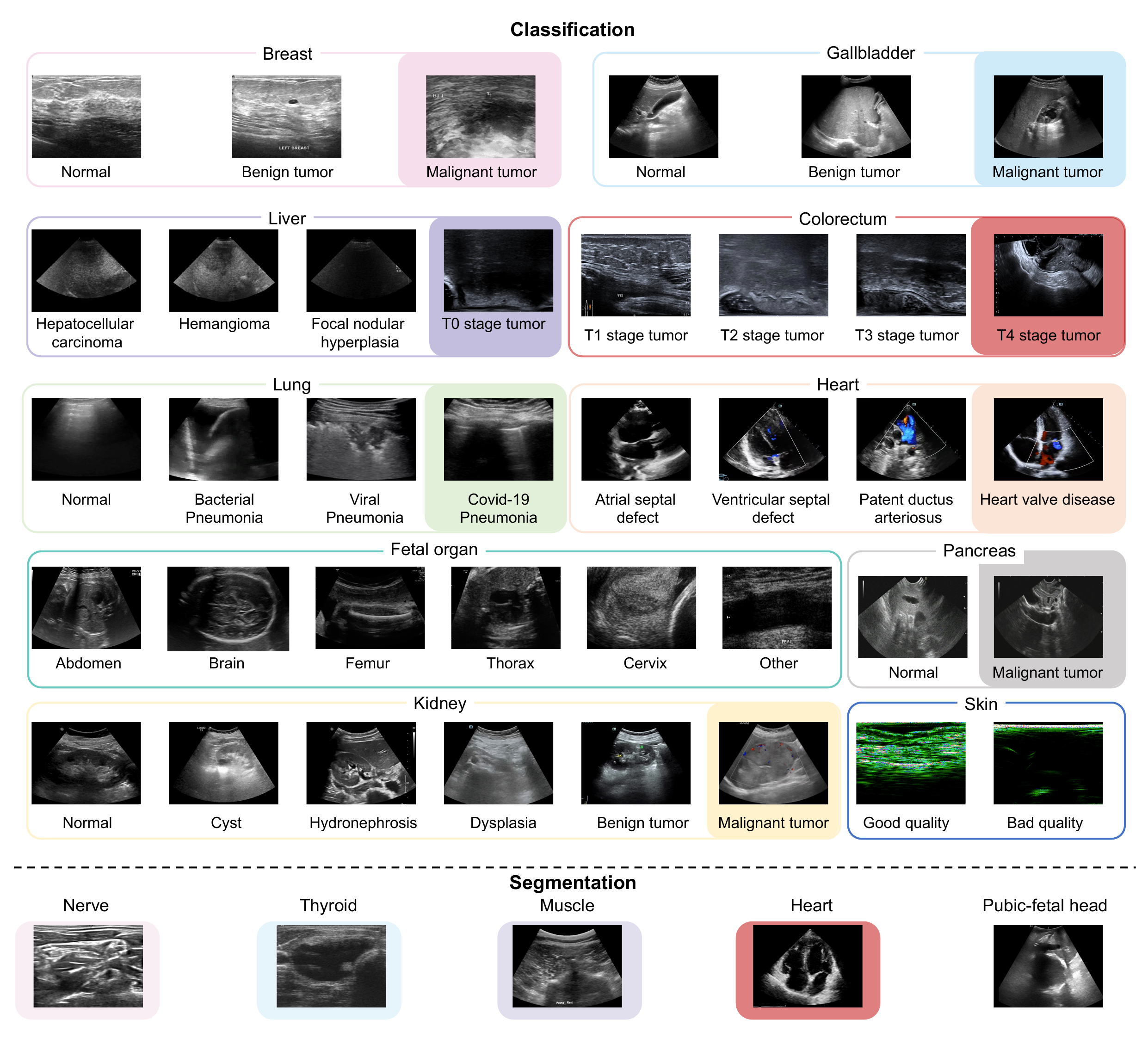}
    \caption{Sample images of each category in the downstream validation dataset. The categories in the color blocks are used to construct the organ-agnostic dataset.}
    \label{fig:example_finetune}
\end{figure}

\begin{figure}[!t]
    \centering
    \includegraphics[width=\linewidth]{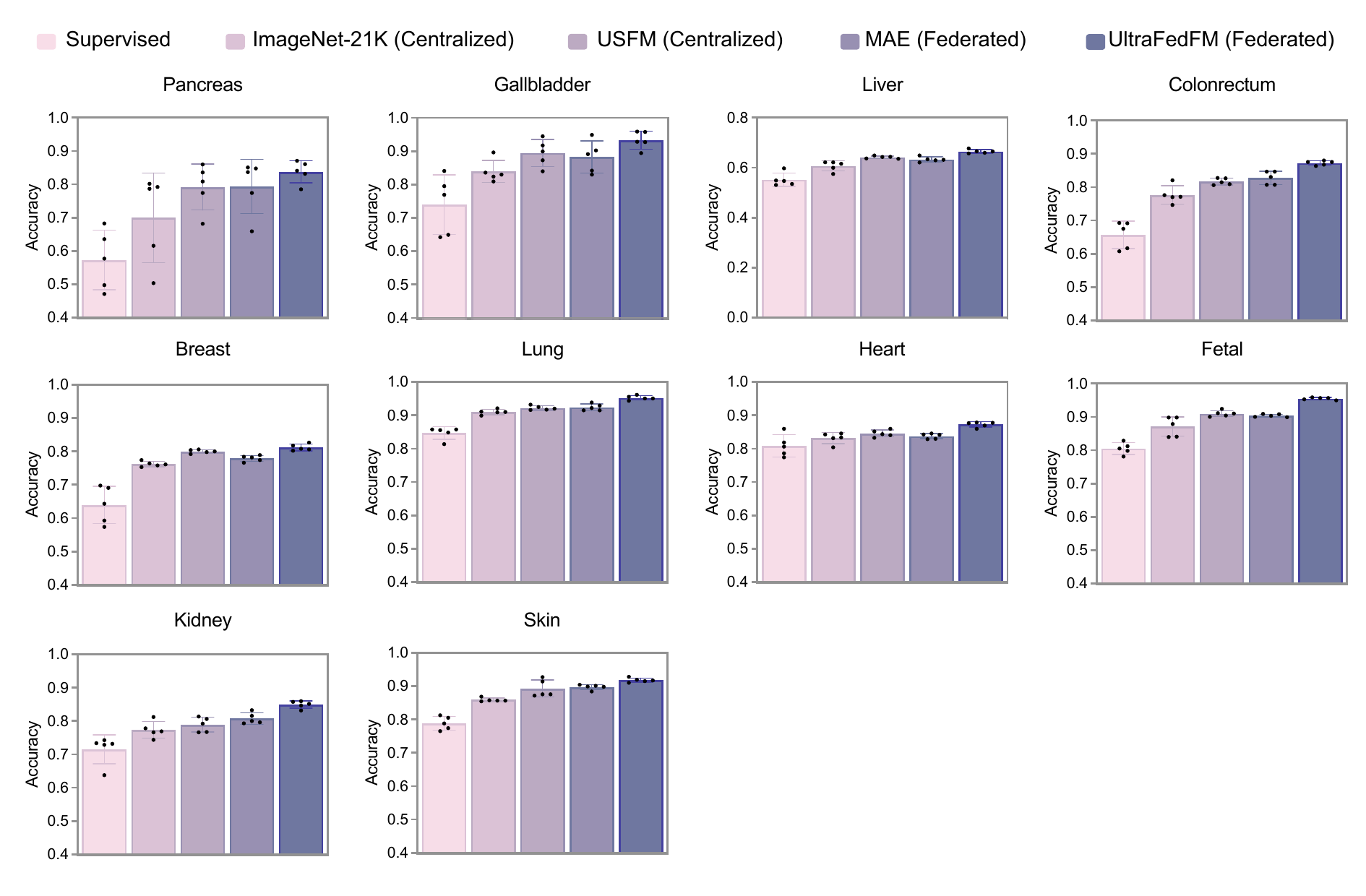}
    \caption{\textbf{Performance (Accuracy) on systemic disease diagnosis.} For each task, we trained the model with 5 different random seeds, determining the shuffling of training data, and evaluated the models on the test set to get 5 replicas. We derived the statistics with the 5 replicas. The error bars show 95\% confidence intervals and the bars’ center represents the mean value of the accuracy. }
    \label{supp:accuracy}
\end{figure}

\begin{figure}[!t]
    \centering
    \includegraphics[width=\linewidth]{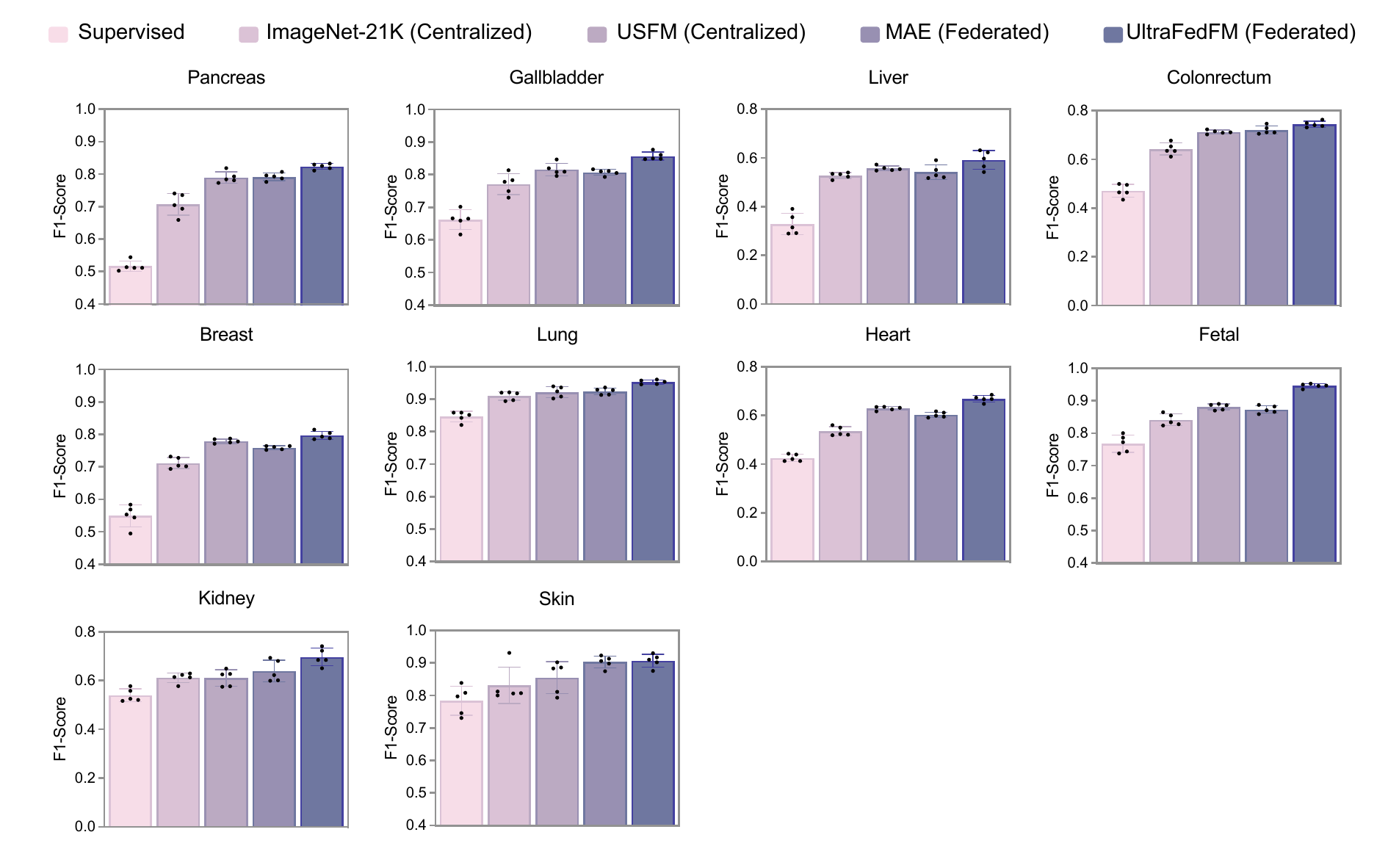}
    \caption{\textbf{Performance (F1-score) on systemic disease diagnosis.} For each task, we trained the model with 5 different random seeds, determining the shuffling of training data, and evaluated the models on the test set to get 5 replicas. We derived the statistics with the 5 replicas. The error bars show 95\% confidence intervals and the bars’ center represents the mean value of the f1-score.}
    \label{supp:f1}
\end{figure}

\begin{figure}[!t]
    \centering
    \includegraphics[width=\linewidth]{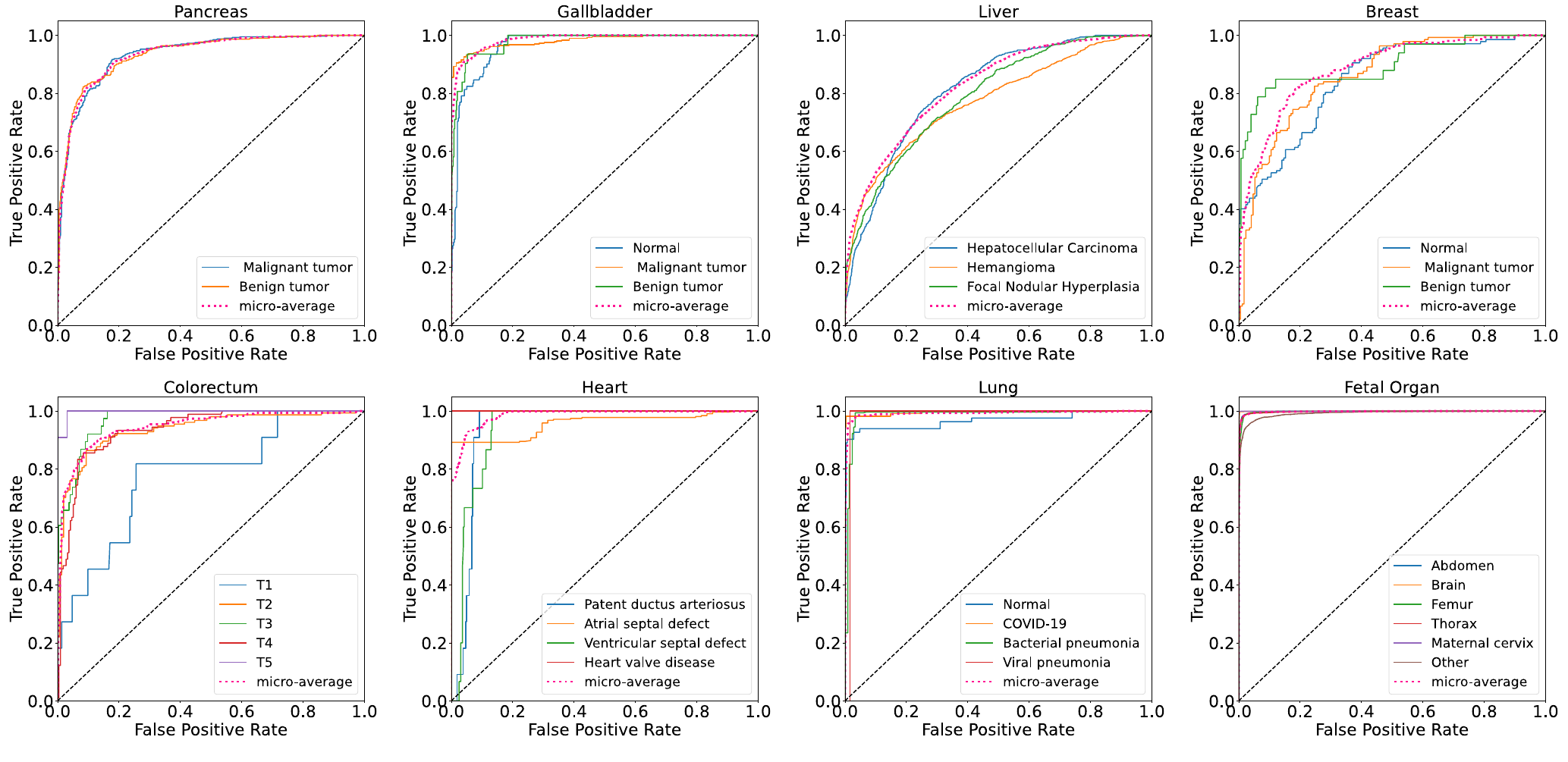}
    \caption{\textbf{ROC of UltraFedFM on each systemic disease diagnosis tasks.}}
    \label{supp:roc}
\end{figure}

\begin{figure}[!t]
    \centering
    \includegraphics[width=\linewidth]{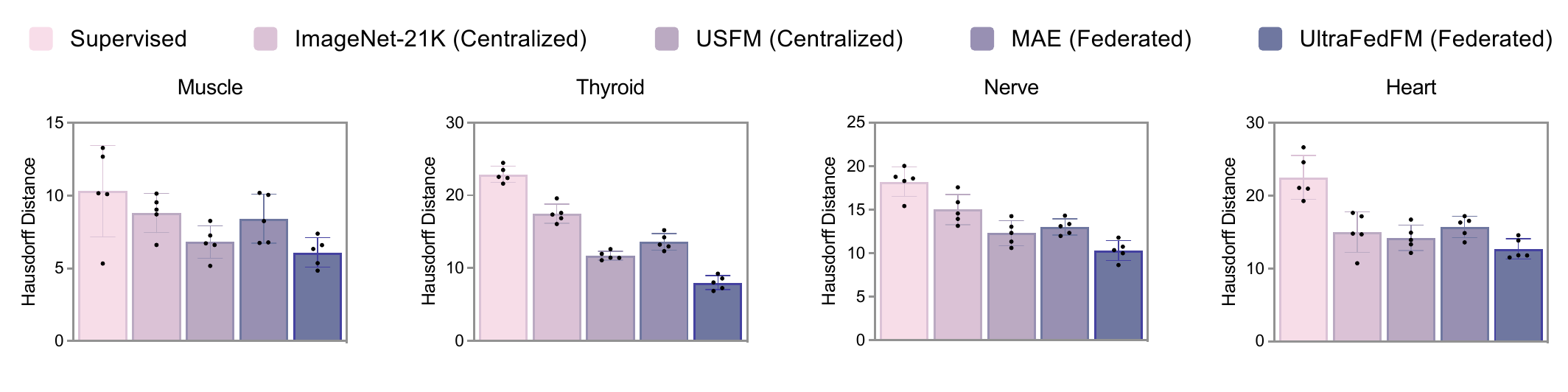}
    \caption{\textbf{Performance (Hausdorff Distance) on organ and lesion segmentation.} For each task, we trained the model with 5 different random seeds, determining the shuffling of training data, and evaluated the models on the test set to get 5 replicas. We derived the statistics with the 5 replicas. The error bars show 95\% confidence intervals and the bars’ center represents the mean value of the f1-score.}
    \label{supp:hd}
\end{figure}

\begin{figure}[!t]
    \centering
    \includegraphics[width=\linewidth]{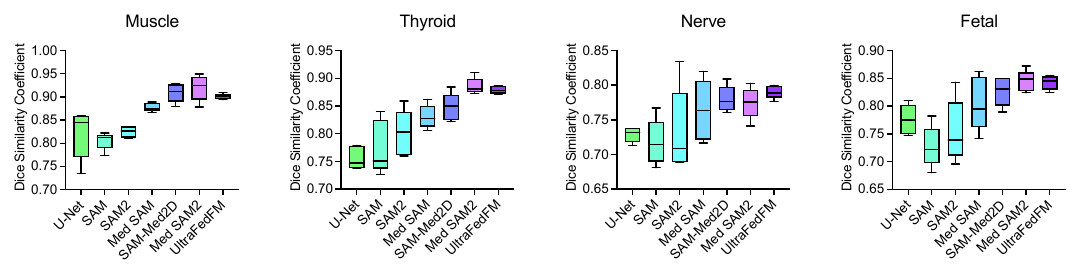}
    \caption{\textbf{Performance (Dice similarity coefficient) comparison of SAM-based models and UltraFedFM on organ and lesion segmentation.} For each task, we trained the model with 5 different random seeds, determining the shuffling of training data, and evaluated the models on the test set to get 5 replicas. We derived the statistics with the 5 replicas. The error bars show 95\% confidence intervals and the bars’ center represents the mean value of the Dice similarity coefficient.}
    \label{supp:sam}
\end{figure}

\end{document}